\documentclass[journal]{new-aiaa}
\usepackage[utf8]{inputenc}
\usepackage{graphicx}
\usepackage{amsmath}
\usepackage[version=4]{mhchem}
\usepackage{siunitx}
\usepackage{longtable,tabularx}
\usepackage{subcaption}
\usepackage{algorithm}  
\usepackage{algorithmicx} 
\usepackage{algpseudocode}
\title{Confidence-Aware Learning Optimal Terminal Guidance via Gaussian Process Regression}

\author{
Han Wang \footnote{Ph.D. Student, School of Astronautics, kingham@buaa.edu.cn.}
and
Donghe Chen \footnote{Ph.D. Student, School of Astronautics, PaoShou@buaa.edu.cn.}
and
Tengjie Zheng \footnote{Ph.D. Student, School of Astronautics, ZhengTengjie@buaa.edu.cn.}

and
Lin Cheng \footnote{\textbf{Corresponding Author}, Associate Professor, School of Astronautics, chenglin5580@buaa.edu.cn. Member AIAA.}
and
Shengping Gong \footnote{Professor, School of Astronautics, gongsp@buaa.edu.cn. Member AIAA.}
}
\affil{Beihang University, 100191 Beijing, China}

\begin{document}

\maketitle

\begin{abstract}
	Modern aerospace guidance systems demand rigorous constraint satisfaction, optimal performance, and computational efficiency. Traditional analytical methods struggle to simultaneously satisfy these requirements. While data driven methods have shown promise in learning optimal guidance strategy, challenges still persist in generating well-distributed optimal dataset and ensuring the reliability and trustworthiness of learned strategies. This paper presents a confidence-aware learning framework that addresses these limitations. First, a region-controllable optimal data generation method is proposed leveraging Hamiltonian state transition matrices, enabling efficient generation of optimal trajectories of specified data distribution. Then, to obtain a lightweight and effective dataset for efficient strategy learning, an error-distribution-smoothing method is incorporated to employ data filtering, which reduces dataset size by almost 90\% while preserving prediction accuracy. To assess the operational domain of the learned strategy, a confidence-aware learning guidance strategy is proposed based on gaussian process regression, achieving constraint satisfaction even beyond training distributions. Numerical simulations validate the effectiveness and reliability of the proposed learning framework in terms of data generation, data filtering and strategy learning. 
\end{abstract}

\section{Introduction}
Terminal guidance systems for aerospace vehicles demand rigorous constraint satisfaction, optimality in energy consumption, and real-time computational efficiency. Although traditional analytical approaches, such as Lyapunov theory \cite{sastry_lyapunov_1999} and Pontryagin's minimum principle \cite{bryson_applied_1975}, have achieved partial success, they struggle to simultaneously satisfy the aforementioned requirements. Recent advances in data-driven methods have emerged as promising alternatives, enabling the direct learning of guidance strategies from optimal trajectory data \cite{chai_real-time_2020,wang_neural-network-based_2024,lu_learning-based_2024,origer_certifying_2024}. However, critical challenges persist: A well-distributed dataset that covers the desired operational envelope often requires extensive sampling, resulting in a heavy computational burden. Few studies pay attention to evaluating when and where the learned strategies can be trusted, posing risks for safety-critical aerospace applications. This paper aims to addresses these issues by developing a confidence-aware learning framework for constrained optimal terminal guidance problems.

For handling nonlinearity and constraints for guidance problems, Lyapunov-based methods offer greater flexibility. Several studies improve the proportional navigation (PN) leveraging Lyapunov theory so that different terminal constraints, such as impact time constraints \cite{jeon_impact-time-control_2016, cho_modified_2016, dong_varying-gain_2023} or impact angle constraints \cite{ratnoo_impact_2008, tekin_switched-gain_2015, ratnoo_impact_2010, erer_indirect_2012, kim_bias-shaping_2013}, are satisfied. Besides, sliding mode control has also been directly used to achieve impact time or impact angle control guidance \cite{harl_impact_2012,kim_lyapunov-based_2015,saleem_lyapunov-based_2016,cho_nonsingular_2016,wang_sliding_2016,yang_three-dimensional_2022,hou_terminal_2019}. However, these studies fail to guarantee the optimality of the guidance strategy, which can be a significant limitation in applications requiring precise performance metrics.

To achieve optimality during the guidance process, Pontryagin's minimum principle has provided a theoretical foundations for constrained optimal guidance problems. In \cite{song_optimal_2004}, Song et al. derives an optimal guidance law with imapct angle constraints by solving a linear quadratic optimal control problem. For arbitrary-order linear missile dynamics, a linear quadratic optimal guidance law with intercept angle constraints is obtained in \cite{shaferman_linear_2008}, and \cite{mishley_linear_2022} further investigates guidance problems with intercept angle constraints and varying speed adversaries. For impact time control guidance problems, a minimum effort guidance law is derived in \cite{merkulov_minimum-effort_2022} by solving a quadratic approximation optimal control problem. However, these achievements are predominantly confined to linear systems or rely on linearization assumptions \cite{cheng_neural-network-based_2024}. The inherent nonlinearity of realistic missile dynamics makes it analytically intractable or computationally prohibitive solve two-point boundary value problems online, which limits its practical applicability.

In recent years, data-driven methods have shown great potential in addressing nonlinear optimal guidance and control problems with satisfying real-time performance \cite{izzo_optimality_2024,qu_experience_2025,zheng_model_2025}. Data distribution and learning method have significant influence on the efficiency and effectiveness of the strategy learning. Traditional shooting methods face extensive computational burden when generating massive optimal trajectories that satisfy terminal constraints. To address this problem, \cite{izzo_real-time_2021} propose a backward generation of optimal examples (BGOE) method, where terminal constraints and optimal conditions are directly integrated in the backward integral process. Utilizing the BGOE method, neural-network-based nonlinear optimal guidance strategy have been investigated to satisfy impact angle constraints and impact time constraints in \cite{cheng_neural-network-based_2024} and \cite{wang_nonlinear_2022}, respectively.  However, the spatial distribution of generated data by the BGOE method remains uncontrollable, affecting the effectiveness and efficiency of supervised learning. Moreover, above neural-network-based methods rarely discuss when the learned strategy is effective, posing potential risks in practical applications. 

In conclusion, critical challenges remain unsolved in controllably generating a well-distributed dataset and rigorous quantifying confidence bounds for the learned strategies. To address these issues, we propose a confidence-aware learning method for constrained optimal terminal guidance. The main contributions of this study lie in three aspects: (1) We propose a state-transition-matrix (STM)-guided optimal data generation method by improve the BGOE method, which achieves controllable generation of optimal trajectories of specified initial state. (2) To further reduce the data size while preserving prediction accuracy, we introduce a error-distribution-smoothing (EDS)-based data filtering method, which reduce the data size by 90\%. (3) To evaluate the confidence of the learned strategy, a gaussian-process-regression (GPR)-based confidence-aware learning method is proposed for approximating the optimal command. Then a confidence-weighted guidance strategy is designed to improve the reliability under low-confidence regions. Simulation results for an impact angle control optimal guidance problem have shown that the proposed confidence-aware learning framework generates a lightweight and well-distributed dataset, and extends the applicability of the learned strategy.

The remainder of this paper is organized as follows: Section II presents the formulation of the impact time control optimal guidance problem and introduce the BGOE method. Section III proposes confidence-aware learning method including data generation, data filtering and strategy learning methods. In Section IV, simulations are conducted to validate the effectiveness of the proposed algorithm. Section V concludes this study.

\section{Problem Formulation and Preliminaries}
\label{sec:Formulation}
In this section, we first present the formulation of the impact-time control guidance problem. To facilitate the subsequent method description and application, we will simplify the problem through nondimensionalization and dimensionality reduction. Then the optimality conditions are derived based on the minimum principle and the backward generation of optimal examples (BGOE) method is introduced, forming the basis for subsequent data generation method.

\subsection{Problem Formulation of Imapct Time Control Optimal Guidance}
\begin{figure}[hbt!]
    \centering
    \includegraphics[width=0.4\textwidth]{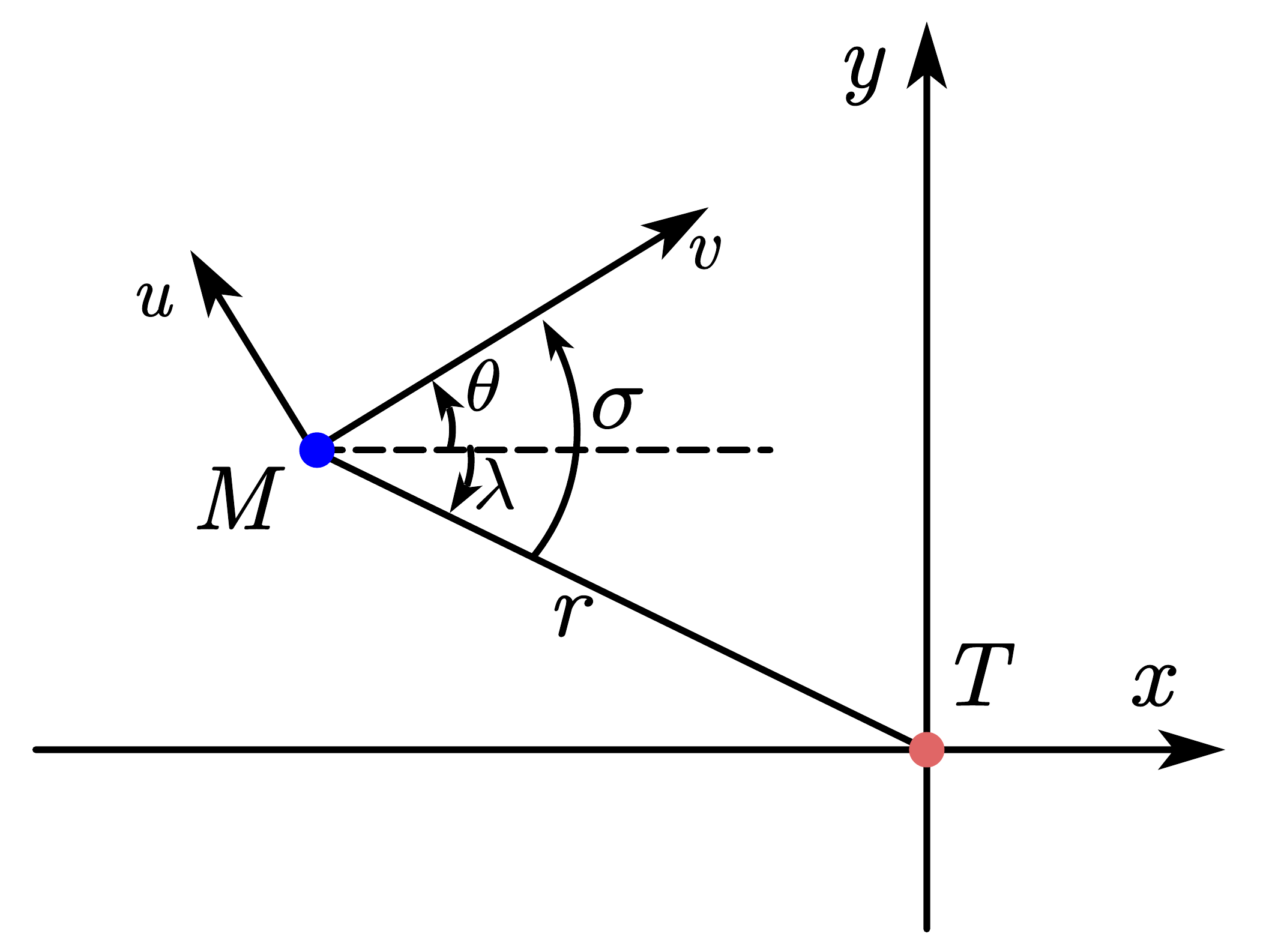}
    \caption{\label{fig: engagement geometry} Planar engagement geometry}
\end{figure}

Consider a terminal guidance scenario where a missile $M$ strikes a stationary target $T$ in a two-dimensional plane. Fig. \ref{fig: engagement geometry} illustrates the engagement geometry. The missile is assumed to maintain a constant speed $v$, with its flight path angle denoted as $\theta$. The relative distance between the missile and the target is denoted by $r$. The line-of-sight (LOS) angle and the heading error angle are represented by $\lambda$ and $\sigma$, respectively. The acceleration of the missile, denoted as $u$, is perpendicular to its velocity.

According to Fig. \ref{fig: engagement geometry}, the engagement dynamics in polar coordinates are governed by
\begin{equation}
    \label{motion equation}
\left\{ \begin{aligned}
	\frac{\text{d}r}{\text{d}t}&=-v\cos \left( \theta -\lambda \right)\\
	\frac{\text{d}\lambda}{\text{d}t}&=-\frac{v\sin \left( \theta -\lambda \right)}{r}\\
	\frac{\text{d}\sigma}{\text{d}t}&=\frac{u}{v}+\frac{v\sin \left( \theta -\lambda \right)}{r}\\
\end{aligned} \right.
\end{equation}

The initial state of the missile is specified as
\begin{equation}
    r(0) = r_0, \lambda(0)=\lambda_0, \sigma(0)=\sigma_0
\end{equation}

For impact-time control guidance, the missile must hit the target at a fixed time $t_f$, leading to terminal constraints expressed as 
\begin{equation}
	\label{terminal constraints}
    r(t_f)=0, \sigma(t_f)=0
\end{equation}

To minimize energy consumption during the guidance process, the performance index is set as
\begin{equation}
    J = \int_{0}^{t_f}\frac{1}{2}u^2\mathrm dt
\end{equation}

Additionally, physical actuator limitations impose a constraint on the acceleration
\begin{equation}
    \left|u\right| \le u_m
\end{equation}

This formulation constitutes a nonlinear optimal control problem. Pontryagin's minimum principle provides theoretical foundations for deriving optimal conditions. To facilitate the following derivation, we will first simplify the problem through nondimensionalization and dimensionality reduction.

\subsection{Nondimensionalization and Dimensionality Reduction of Optimal Trajectories}

\begin{figure}[hbt!]
    \centering
    \includegraphics[width=0.4\textwidth]{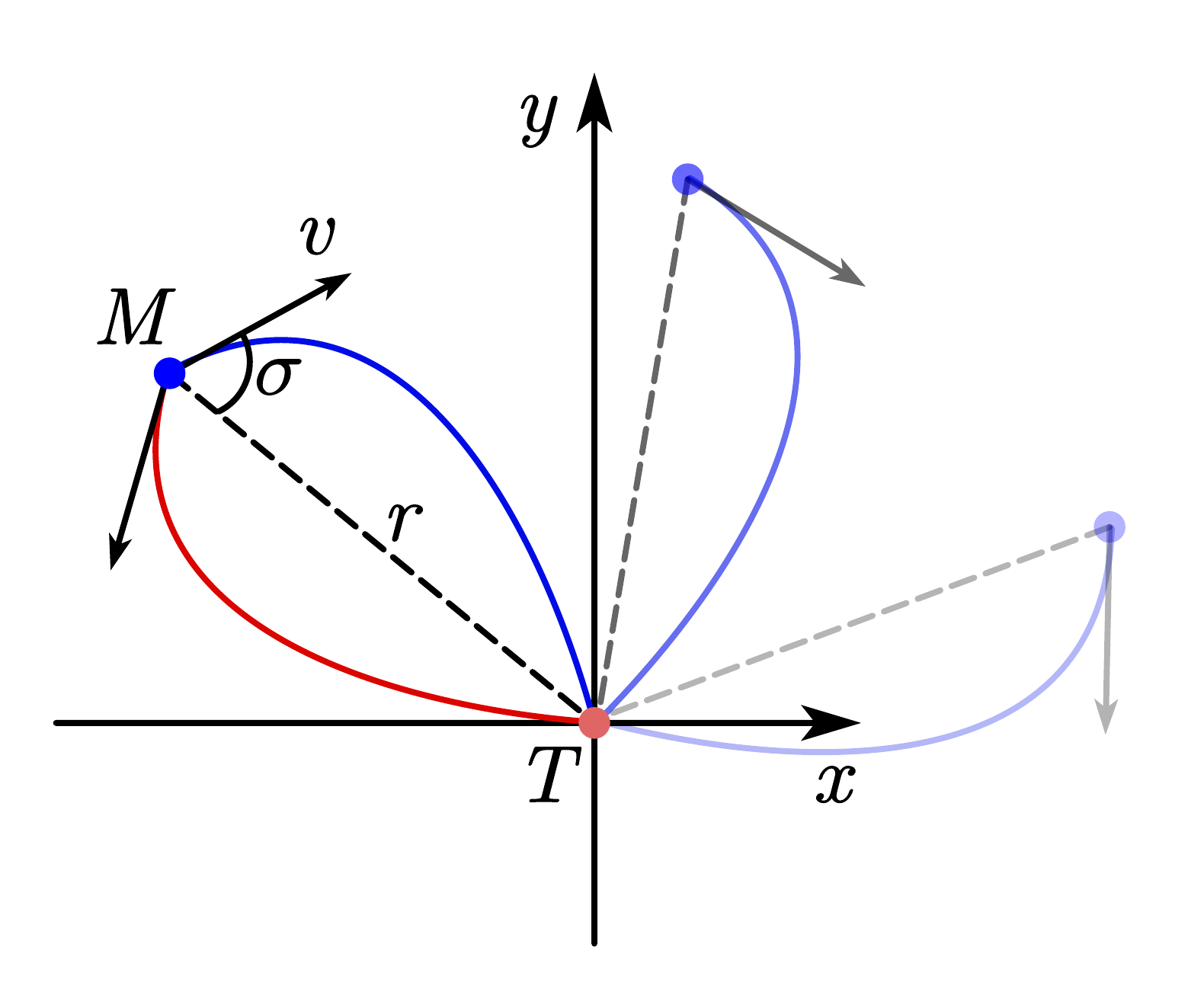}
    \caption{\label{fig: Dimensionality reduction} Dimensionality reduction by symmetry}
\end{figure}

To simplify the problem, we first introduce nondimensionalization and dimensionality reduction techniques in this subsection. Setting distance and time references as $R=r_0$ and $T=r_0/v$, we can derive the acceleration reference as $U=v^2/r$. To avoid notational abuse, we continue using the same symbols for the nondimensionalized values. The nondimensionalized system equation can be rewritten as
\begin{equation}
\left\{ \begin{aligned}
	\frac{\text{d}r}{\text{d}t}&=-\cos \left( \theta -\lambda \right)\\
	\frac{\text{d}\lambda}{\text{d}t}&=-\frac{\sin \left( \theta -\lambda \right)}{r}\\
	\frac{\text{d}\sigma}{\text{d}t}&=u+\frac{\sin \left( \theta -\lambda \right)}{r}\\
\end{aligned} \right. 
\end{equation}

Fig. \ref{fig: Dimensionality reduction} illustrates the symmetry of optimal trajectories. It can be observed that the dark red and dark blue trajectory are axially symmetric with respect to the LOS direction. All blue trajectories can be obtained by rotating one of them. Utilizing rotational symmetry and axial symmetry about the LOS direction, we can reduce the dimensionality of the original system equation by choosing $(r,\sigma)$ as the system state. The reduced system equation can be expressed as
\begin{equation}
	\label{reduced system}
\left\{ \begin{aligned}
	\frac{\text{d}r}{\text{d}t}&=-\cos \sigma\\
	\frac{\text{d}\sigma}{\text{d}t}&=u+\frac{\sin \sigma}{r}\\
\end{aligned} \right. 
\end{equation}

The above simplification facilitates the subsequent derivation and reduces the data requirements. Next we will apply Pontryagin's minimum principle to the reduced system, from which the corresponding optimality conditions are derived.

\subsection{Optimal Conditions and Introduction of the BGOE method}

According to the minimum principle, the Hamiltonian function is defined as
\begin{equation}
H=\frac{1}{2}u^2-p_r\cos \sigma +p_{\sigma}\left( u+\frac{\sin \sigma}{r} \right)
\end{equation}
where $p_r$ and $p_\sigma$ denote the costate variables associated with the relative distance and heading error angle, respectively.

The time derivatives of costates are governed by
\begin{equation}
\left\{ \begin{aligned}
	\dot{p}_r&=-\frac{\partial H}{\partial r}=\frac{p_{\sigma}\sin \sigma}{r^2}\\
	\dot{p}_{\sigma}&=-\frac{\partial H}{\partial \sigma}=-p_r\sin \sigma -\frac{p_{\sigma}\cos \sigma}{r}\\
\end{aligned} \right. 
\end{equation}

According to the extremum condition, the optimal control command can be derived by minimizing the Hamiltonian function, which is expressed as
\begin{equation}
	\label{optimal u}
u^*=\underset{u\in U}{\text{arg}\min}\ H=\left\{ \begin{matrix}
	u_m,&		p_{\sigma}>u_m\\
	p_{\sigma},&		-u_m\le p_{\sigma}\le u_m\\
	-u_m,&		p_{\sigma}<-u_m\\
\end{matrix} \right. 
\end{equation}

Since there are both terminal constraints on the relative distance $r$ and the heading error angle $\sigma$, terminal conditions for costates are
\begin{equation}
	\label{transversal condition}
\left\{ \begin{array}{l}
	p_r\left( t_f \right) =p_{rf}\\
	p_{\sigma}\left( t_f \right) =p_{\sigma f}\\
\end{array} \right.
\end{equation}

Equations \eqref{reduced system}$\sim$\eqref{transversal condition} constitute a classical two-point boundary value problem, which is hard to solve analytically. While conventional numerical approaches (e.g., shooting methods) incur substantial computational costs, the BGOE method enables rapid generation of optimal trajectories. We begin by introducing the BGOE framework in the impact time optimal guidance problem. Subsequent analysis reveals inherent limitations of the standard BGOE approach, thereby motivating our proposed enhancements.

The BGOE method first defines a parameterized system leveraging the original Hamiltonian system and above derived optimal conditions. The parameterized system is expressed as
\begin{equation}
	\label{parameterized system}
\left\{ \begin{aligned}
	\frac{\text{d}\tilde{r}}{\text{d}t}&=\cos \tilde{\sigma}\\
	\frac{\text{d}\tilde{\sigma}}{\text{d}t}&=-\left( \tilde{u}+\frac{\sin \tilde{\sigma}}{\tilde{r}} \right)\\
	\frac{\text{d}\tilde{p}_r}{\text{d}t}&=-\frac{\tilde{p}_{\sigma}\sin \tilde{\sigma}}{\tilde{r}^2}\\
	\frac{\text{d}\tilde{p}_{\sigma}}{\text{d}t}&=\tilde{p}_r\sin \tilde{\sigma}+\frac{\tilde{p}_{\sigma}\cos \tilde{\sigma}}{\tilde{r}}\\
\end{aligned} \right. 
\end{equation}
where $\tilde u$ represents the optimal command for the parameterized system, expressed as
\begin{equation}
	\tilde{u}=\left\{ \begin{matrix}
		u_m,&		\tilde{p}_{\sigma}>u_m\\
		\tilde{p}_{\sigma},&		-u_m\le \tilde{p}_{\sigma}\le u_m\\
		-u_m,&		\tilde{p}_{\sigma}<-u_m\\
	\end{matrix} \right. 
\end{equation}

Note that this parameterized system moves in the opposite direction of the original Hamiltonian system, and optimal conditions \eqref{reduced system}$\sim$\eqref{optimal u} are satisfied along its trajectories. Therefore, by choosing terminal conditions of the original system \eqref{terminal constraints} and \eqref{transversal condition} as the initial state of the parameterized system, one can directly obtain the optimal trajectory by integrating the parameterized system for desired time constraint $t_f$. 

Although the BGOE method enables rapid generation of optimal trajectories, it has to determine terminal costates \eqref{transversal condition} to start the backward integral process. The absence of rigorous selection criteria for terminal costates lead to incomplete and uncontrollable distribution of generated optimal data. Furthermore, when deploying neural-network-based learning methods \cite{wang_neural-network-based_2024,cheng_neural-network-based_2024} on such irregular datasets, the confidence risk for practical applications of learned strategies becomes more severe. These dual challenges motivate the development of our confidence-aware learning framework presented in the next section, where a lightweight and well-distributed dataset is generated and the confidence of the learned strategy is evaluated.

\section{Confidence-Aware Learning for Constrained Optimal Terminal Guidance}
In this section, we first present a state-transition-matrix-guided optimal data generation method to obtain a region-controllable dataset. Then we propose a error-distribution-smoothing-based data filtering method to reduce data quantity while preserving prediction accuracy. To reduce application risks, we propose a GPR-based confidence-aware learning method, enabling confidence evaluation and improving the reliability of the learned strategy.

\subsection{State-Transition-Matrix-Guided Optimal Data Generation}
To address the drawback of incomplete and uncontrollable data generation in the BGOE method, we proposed a region management technique by introducing the state transition matrix of the Hamiltonian system, enabling precise control over the spatial and temporal distribution of generated trajectories.

To derive the rule of choosing terminal costates, we first analyze how terminal costates influence the generated state. Let $\boldsymbol{Z}=\left[ \boldsymbol{X}^T,\boldsymbol{P}^T \right] ^T$, where $\boldsymbol{X}=\left[ \tilde{r},\tilde{\sigma} \right] ^T$ represents the state, and $\boldsymbol{P}=\left[ \tilde{p}_r,\tilde{p}_{\sigma} \right] ^T$ represent the costate. The parameterized system can be rewritten as $\boldsymbol{\dot{Z}}=\boldsymbol{F}\left( \boldsymbol{Z} \right) $. 

Define the state transition matrix as $\boldsymbol{\Phi }\left( t_f,t_0 \right)$, which represents the effect of a small change $\delta \boldsymbol{Z}\left( t_0 \right)$ on the state $\delta \boldsymbol{Z}\left( t_f \right)$ after a period from $t_0$ to $t_f$. Therefore, it holds that
\begin{equation}
    \label{STM}
\delta \boldsymbol{Z}\left( t_f \right) =\boldsymbol{\Phi }\left( t_f,t_0 \right) \delta \boldsymbol{Z}\left( t_0 \right) , \boldsymbol{\Phi }\left( t_i,t_i \right) =\boldsymbol{I}
\end{equation}

It can be derived that the state transition matrix satisfies the following differential equation.
\begin{equation}
	\label{STM DF}
\boldsymbol{\dot{\Phi}}\left( t,t_0 \right) =\frac{\text{d}}{\text{d}t}\left( \frac{\partial \boldsymbol{Z}\left( t \right)}{\partial \boldsymbol{Z}\left( t_0 \right)} \right) =\frac{\partial}{\partial \boldsymbol{Z}\left( t_0 \right)}\left( \frac{\text{d}\boldsymbol{Z}\left( t \right)}{\text{d}t} \right) =\frac{\partial \boldsymbol{F}\left( \boldsymbol{Z}\left( t \right) \right)}{\partial \boldsymbol{Z}\left( t \right)}\frac{\partial \boldsymbol{Z}\left( t \right)}{\partial \boldsymbol{Z}\left( t_0 \right)}=\boldsymbol{F}_{\boldsymbol{Z}}\boldsymbol{\Phi }\left( t,t_0 \right) 
\end{equation}
where $\boldsymbol{F}_{\boldsymbol{Z}}$ denotes the Jacobian matrix of the parameterized system equation.

By integrating the parametric system (\ref{parameterized system}) and the differential equation of the state transition matrix (\ref{STM DF}) from time $t_0$ to $t_f$, we can obtain the state transition matrix $\boldsymbol{\Phi }\left( t_f, t_0 \right)$. It satisfies 
\begin{equation}
	\label{xp-x}
\begin{aligned}
	\delta \boldsymbol{X}\left( t_f \right) &=\frac{\partial \boldsymbol{X}\left( t_f \right)}{\partial \boldsymbol{Z}\left( t_f \right)}\boldsymbol{\Phi }\left( t_f,t_0 \right) \delta \boldsymbol{Z}\left( t_0 \right)\\
	&=\left[ \begin{matrix}
	0&		\boldsymbol{I}\\
\end{matrix} \right] \boldsymbol{\Phi }\left( t_f,t_0 \right) \left[ \begin{array}{c}
	\delta \boldsymbol{X}\left( t_0 \right)\\
	\delta \boldsymbol{P}\left( t_0 \right)\\
\end{array} \right]\\
\end{aligned}
\end{equation}

Note that the parameterized system moves backward compared with the original Hamiltonian system. Equation \eqref{xp-x} gives the relationship that variations of terminal states $\delta \boldsymbol{X}\left( t_0 \right)$ and costates $\delta \boldsymbol{P}\left( t_0 \right)$ influence on the variation of generated states $\delta \boldsymbol{X}\left( t_f \right)$.

Given that the guidance problem has fixed terminal constraints, which means the parameterized system satisfies $\delta \boldsymbol{X}\left( t_0 \right) =0$, we can derive the relationship between variations of terminal costates $\delta \boldsymbol{P}\left( t_0 \right)$ and generated states $\delta \boldsymbol{X}\left( t_f \right)$ as
\begin{equation}
    \label{BGOE-ITR-px0pxf}
\frac{\partial \boldsymbol{X}\left( t_f \right)}{\partial \boldsymbol{P}\left( t_0 \right)}=\left[ \begin{matrix}
	0&		\boldsymbol{I}\\
\end{matrix} \right] \Phi \left( t_f,t_0 \right) \left[ \begin{array}{c}
	0\\
	\boldsymbol{I}\\
\end{array} \right]
\end{equation}

Based on the above equation, we can manipulate the generated optimal data $\boldsymbol X(t_f)$ by changing the selected terminal costate $\boldsymbol P(t_0)=[p_{rf}, p_{\sigma f}]^T$. Through above analysis, we provide a principle of setting the terminal costate when a specific change $\delta \boldsymbol{X}_f$ for the generated state is desired.
\begin{equation}
    \label{delta_p01}
\delta \boldsymbol{P}_0=\left( \frac{\partial \boldsymbol{X}\left( t_f \right)}{\partial \boldsymbol{P}\left( t_0 \right)} \right) ^{-1}\delta \boldsymbol{X}_f
\end{equation}

In addition, for generating optimal trajectories with a specific change of terminal time constraint $\delta t_f$, the change of terminal costate can also be determined as 
\begin{equation}
    \label{delta_p02}
\delta \boldsymbol{P}_0=\left( \frac{\partial \boldsymbol{X}\left( t_f \right)}{\partial \boldsymbol{P}\left( t_0 \right)} \right) ^{-1}\frac{\text{d}\boldsymbol{X}\left( t_f \right)}{\text{d}t}\delta t_f
\end{equation}

Using equations (\ref{delta_p01}) and (\ref{delta_p02}), we can adjust the selection of costates during the integration of the parameterized system to obtain the desired state $\boldsymbol{X}_{d}$ and terminal time constraint $t_{d}$, which can be concluded as
\begin{equation}
	\label{costate adjusting}
\boldsymbol{P}_{0,i+1}=\boldsymbol{P}_{0,i}+\left( \frac{\partial \boldsymbol{X}\left( t_{d,i} \right)}{\partial \boldsymbol{P}\left( t_0 \right)} \right) ^{-1}\left( \boldsymbol{X}_{{d,}i+1}-\boldsymbol{X}_{{d},i} \right) +\left( \frac{\partial \boldsymbol{X}\left( t_{d,i} \right)}{\partial \boldsymbol{P}\left( t_0 \right)} \right) ^{-1}\frac{\text{d}\boldsymbol{X}\left( t_{d,i} \right)}{\text{d}t}\left( t_{d,i+1}-t_{d,i} \right) 
\end{equation}
where $\boldsymbol{P}_{0,i}$, $\boldsymbol{X}_{{d},i}$ and $t_{d,i}$ respectively denotes the selected costate, the generated optimal state and the terminal time constraint at previous integral process, $\boldsymbol{X}_{{{d},i+1}}$ and $t_{d,i+1}$ denotes the desired state and terminal time constraint, $\boldsymbol{P}_{0,i+1}$ denotes the selected costate at next integral process.

\floatname{algorithm}{Algorithm}
\begin{algorithm*}[!h]
	
  \caption{\label{Algorithm1} State-transition-matrix-guided backward generation of optimal examples}
  \begin{algorithmic}[1]
	\Statex \textbf{Input:} Desired initial state region $r\in \left[ r_{\min},r_{\max} \right] ,\sigma \in \left[ \sigma _{\min},\sigma _{\max} \right] ,t_f\in \left[ t_{\min},t_{\max} \right]$, State step $\Delta r, \Delta \sigma, \Delta t_f$, Acceleration constraint $u_m$.

	\Statex \textbf{Output:} Optimal dataset $\mathcal{D}=\left\{ \left( r,\sigma ,t_{\text{go}},u \right) \right\}$

	\State Initialize: Set dataset $\mathcal D = \emptyset$ and the desired initial state $ r_{d,i}=r_{\min}, \sigma_{d,i}=\sigma_{\min}, t_{d,i}={t_{\min}}$. Calculate the corresponding terminal costate $\boldsymbol{P}_{0,i}=\left[ p_{rf},p_{\sigma f} \right] ^T$ w.r.t. the desired initial state.

	\State Integrate the parameterized system (\ref{parameterized system}) and (\ref{STM DF}) for time $t_{d,i}$, store the generated optimal data $(r, \sigma, t_{\text{go}}, u)$ into $\mathcal D$ and obtain the state transition matrix $\boldsymbol{\Phi }\left(t_{d,i},0 \right)$.

	\For {each next desired initial state ($r_{d,i+1}, \sigma_{d, i+1}, t_{d,i+1})$ that traverses the state region with setting steps} 
		\State Adjust the terminal constates by (\ref{costate adjusting}) and obtain $\boldsymbol{P}_{0, i+1}$.
		\State Set $r_{d,i}\gets r_{d,i+1},\sigma _{d,i}\gets \sigma _{d,i+1},t_{d,i}\gets t_{d,i+1},\boldsymbol{P}_{0,i}\gets \boldsymbol{P}_{0,i+1}$.
		\State Integrate (\ref{parameterized system}) and (\ref{STM DF}) for time $t_{d,i}$, store optimal data and obtain the state transition matrix $\boldsymbol{\Phi }\left(t_{d,i},0 \right)$.
	\EndFor

	\State End the algorithm. Store optimal dataset $\mathcal D$.
  \end{algorithmic}
\end{algorithm*}

The procedure to generate optimal data within a predefined region is concluded as Algorithm \ref{Algorithm1}. By traversing the desired state region leveraging equation (\ref{costate adjusting}), the desired optimal data can be obtained. For the legibility of the pseudo-code, we omit the details of setting desired initial states that traverses the state region, which is done by walking a serpentine route through the state region.

This subsection presents a region-controllable data generation method that leverages the state transition matrix to adjust terminal costates, enabling the generation of optimal trajectories within predefined regions. Although the proposed method has obtained optimal data in specific regions, due to the requirements of integration accuracy, the data density in the temporal dimension is significantly higher than that in the spatial dimension. This irregular data distribution is detrimental to subsequent learning processes. Therefore, next we will introduce a data filtering technique to further reduce sample size, adjust the data distribution, while preserving learning accuracy.

\subsection{Error Distribution Smoothing based Data Filtering}
Given the large volume and irregular spatio-temporal distribution of generated optimal data, this section introduces a data filtering method based on error distribution smoothing (EDS) \cite{chen_error_2025} to reduce the dataset size while preserving the prediction accuracy. The goal is to retain critical data samples in regions where the function exhibits high complexity while removing redundant samples in simpler regions.

Let the input vector $\boldsymbol{\psi }=\left[ r,\sigma ,t_f \right] ^T$ encapsulate the system state and terminal time constraint, with corresponding optimal control $u$ as the output. The entire dataset is denoted as $\mathcal D = \{\boldsymbol{\Psi}, \boldsymbol{U}\}$, where $\boldsymbol{\Psi}=\{\boldsymbol{\psi}_1, \boldsymbol{\psi}_2,\cdots, \boldsymbol{\psi}_N\}$ and $\boldsymbol{U}=\{u_1, u_2, \cdots, u_N\}$.

The core idea of the EDS method is to select a minimal subset of data points that maintain the required learning accuracy for representing the original dataset. To quantify the balance of function complexity and sample size in a subset $\Omega$, a metric called the Complexity-to-Density Ratio (CDR) is defined as
\begin{equation}
\rho \left( \Omega ,\mathcal{D} \right) =\frac{g_c\left( \Omega \right)}{\left| \Omega \right|/g_s\left( \Omega \right)}=\frac{g_c\left( \Omega \right) \cdot g_s\left( \Omega \right)}{\left| \Omega\right|}
\end{equation}
where $\left|\Omega\right|$ denotes the number of samples in the subset. 

$g_c\left( \Omega \right)$ represents the maximum Frobenius norm of the Hessian matrix, which measures the complexity of function relationship between inputs and outputs.
\begin{equation}
	g_c\left( \Omega \right) =\underset{\boldsymbol{\psi }\in \Omega}{\max}\lVert \boldsymbol{H}\left( \boldsymbol{\psi } \right) \rVert _F
\end{equation}

$g_s\left( \Omega \right)$ denotes the spatial size of the dataset, which is defined as the maximum squared Euclidean distance between any two points in $\Omega$.
\begin{equation}
	g_s(\Omega) = \max_{\boldsymbol{x}_1, \boldsymbol{x}_2 \in \Omega} \| \boldsymbol{x}_1 - \boldsymbol{x}_2 \|_2^2
\end{equation}

To evaluate the distribution of Complexity-to-Density Ratio (CDR) across all regions $\mathcal{F}=\{\Omega_i\}_{i=1}^k$, define the Log-CDR distribution as $I(\mathcal{F},\mathcal{D})\sim\mathcal{N}(\mu_\rho,\sigma_\rho^2)$, 
where 
\begin{equation}
\begin{aligned}
	\mu_\rho&=\frac{1}{k}\sum_{j=1}^k{\ln}\left( \rho \left( \Omega _j,\mathcal{D} \right) \right)\\
	\sigma_\rho^2&=\frac{1}{k-1}\sum_{j=1}^k{\left( \ln \left( \rho \left( \Omega _j,\mathcal{D} \right) \right) -\hat{\mu} \right)}^2\\
\end{aligned}
\end{equation}

Through above analysis, the data filtering problem can be formulated as follows. It aims to minimize the sample size of the selected subset with the complexity-to-density ratio (CDR) less than a certain level, which assures there have enough samples in regions of high function complexity. An optimization problem is defined as (\ref{data filtering problem}) to select the smallest possible subset while maintaining data distribution balance.
\begin{equation}
	\label{data filtering problem}
\begin{aligned}
	&\underset{\mathcal{D}_R\subseteq \mathcal{D}}{\min}|\mathcal{D}_R|\\
	&\text{subject\,\,to\quad }\mu _{\rho}+z\sigma _{\rho}\leq \varepsilon \\
\end{aligned}
\end{equation}
where $z$ is the standard deviation coefficient, $\varepsilon$ is a given threshold that trade off between sample size and representation accuracy.

By applying the error distribution smoothing algorithm in \cite{chen_error_2025}, the optimization problem can be approximately solved, resulting in a filtered subdataset $\mathcal{D}_R$ that retains critical information while significantly reducing data volume. This filtered dataset facilitates more efficient learning for the guidance strategy in subsequent steps.

\subsection{Gaussian Process Regression based Confidence-Aware Strategy Learning}
Current learning methods for optimal datasets primarily rely on deep neural networks that are directly applied to guidance processes. However, neural networks inherently lack the ability to evaluate the confidence of their predictions, thereby posing risks in practical applications. This section proposes a gaussian process regression (GPR)-based method to achieve confidence-aware learning for optimal strategies. The reliability of the learned strategy beyond the training set range is further improved by integrating it with an analytical guidance law according to confidence levels.

The relationship between the optimal guidance command and the system state can be expressed as
\begin{equation}
u_i=f\left( \boldsymbol{\psi}_i \right)
\end{equation}

The GPR method assumes that the function $f\left( \boldsymbol{\psi} \right)$ follows a gaussian process prior
\begin{equation}
f\left( \boldsymbol{\psi } \right) \sim \mathcal{G}\mathcal{P}\left( \mu _{\boldsymbol{w}_1}\left( \boldsymbol{\psi } \right) ,k_{\boldsymbol{w}_2}\left( \boldsymbol{\psi ,\psi '} \right) \right)
\end{equation}
where $\mu _{\boldsymbol{w}_1}\left( \boldsymbol{\psi } \right) $ is the mean function, $k_{\boldsymbol{w}_2}\left( \boldsymbol{\psi ,\psi '} \right) $ is the prior covariance function, and $\boldsymbol w_1, \boldsymbol w_2$ are hyperparameters to be optimized.

According to the gaussian process regression method \cite{gardner_gpytorch_2021}, in the training stage, the hyperparameters $\boldsymbol w_1, \boldsymbol w_2$ are determined by gradient descent to maximize the log marginal likelihood function given the dataset $\mathcal D = \{\boldsymbol{\Psi}, \boldsymbol{U}\}$.
\begin{equation}
\log p\left( \left. \boldsymbol{U} \right|\boldsymbol{\Psi ,w}_1,\boldsymbol{w}_2 \right) =-\frac{1}{2}\left( \boldsymbol{U}-\boldsymbol{m}\left( \boldsymbol{\Psi } \right) \right) ^T\boldsymbol{K}\left( \boldsymbol{\Psi ,\Psi } \right) ^{-1}\left( \boldsymbol{U}-\boldsymbol{m}\left( \boldsymbol{\Psi } \right) \right) -\frac{1}{2}\log \left| \boldsymbol{K}\left( \boldsymbol{\Psi ,\Psi } \right) \right|-\frac{n}{2}\log 2\pi 
\end{equation}
where 
\begin{equation}
	\boldsymbol{m}\left( \boldsymbol{\Psi } \right) =\left[ \mu _{\boldsymbol{w}_1}\left( \boldsymbol{\psi }_1 \right) ,\mu _{\boldsymbol{w}_1}\left( \boldsymbol{\psi }_2 \right) ,\cdots ,\mu _{\boldsymbol{w}_1}\left( \boldsymbol{\psi }_n \right) \right] ^T
\end{equation}
 \begin{equation}
	\boldsymbol{K}\left( \boldsymbol{\Psi ,\Psi } \right) =\left[ \begin{matrix}
		k_{\boldsymbol{w}_2}\left( \boldsymbol{\psi }_1,\boldsymbol{\psi }_1 \right)&		\cdots&		k_{\boldsymbol{w}_2}\left( \boldsymbol{\psi }_1,\boldsymbol{\psi }_n \right)\\
		\vdots&		\ddots&		\vdots\\
		k_{\boldsymbol{w}_2}\left( \boldsymbol{\psi }_n,\boldsymbol{\psi }_1 \right)&		\cdots&		k_{\boldsymbol{w}_2}\left( \boldsymbol{\psi }_n,\boldsymbol{\psi }_n \right)\\
	\end{matrix} \right] 
 \end{equation}

 Given the gaussian process prior, the test input and output pair $(\boldsymbol{\psi}_*,u_*)$ and training data $\{\boldsymbol{\Psi},\boldsymbol{U}\}$ form a joint gaussian distribution
\begin{equation}
\left[ \begin{array}{c}
	\boldsymbol{U}\\
	u_*\\
\end{array} \right] \sim \mathcal{N}\left( \left[ \begin{array}{c}
	\boldsymbol{m}\left( \boldsymbol{\Psi } \right)\\
	\mu _{\boldsymbol{w}_1}\left( \boldsymbol{\psi }_* \right)\\
\end{array} \right] ,\left[ \begin{matrix}
	\boldsymbol{K}\left( \boldsymbol{\Psi ,\Psi } \right)&		\boldsymbol{K}\left( \boldsymbol{\Psi ,\psi }_* \right)\\
	\boldsymbol{K}\left( \boldsymbol{\psi }_*,\boldsymbol{\Psi } \right)&		k_{\boldsymbol{w}_2}\left( \boldsymbol{\psi }_*,\boldsymbol{\psi }_* \right)\\
\end{matrix} \right] \right)
\end{equation}

Based on the conditional gaussian distribution, the posterior distribution of the predicted output $u_*$ given the test input $\psi_*$ satisfies
\begin{equation}
	\label{posterior distribution}
\left. u_* \right|\boldsymbol{\Psi ,U,\psi }_*\sim \mathcal{N}\left( \mu _*,\sigma^2_* \right)
\end{equation}
where the mean of predicted output is
\begin{equation}
	\label{mean}
\mu _*=\mu _{\boldsymbol{w}_1}\left( \boldsymbol{\psi }_* \right) +\boldsymbol{K}\left( \boldsymbol{\psi }_*,\boldsymbol{\Psi } \right) \boldsymbol{K}\left( \boldsymbol{\Psi ,\Psi } \right) ^{-1}\left( \boldsymbol{U}-\boldsymbol{m}\left( \boldsymbol{\Psi } \right) \right)
\end{equation}
and the corresponding variance is
\begin{equation}
	\label{variance}
\sigma^2_*=k_{\boldsymbol{w}_2}\left( \boldsymbol{\psi }_*,\boldsymbol{\psi }_* \right) -\boldsymbol{K}\left( \boldsymbol{\psi }_*,\boldsymbol{\Psi } \right) \boldsymbol{K}\left( \boldsymbol{\Psi ,\Psi } \right) ^{-1}\boldsymbol{K}\left( \boldsymbol{\Psi ,\psi }_* \right)
\end{equation}

Equations (\ref{mean}) and (\ref{variance}) provide the predictive mean and variance for the optimal guidance command using the GPR method. Unlike neural network-based strategies, the GPR method gives the optimal command as a gaussian distribution, where the variance evaluates the confidence of the prediction. When the predictive variance is large, the confidence of the predictive guidance command is low. To enhance the reliability of the learned guidance strategy beyond the training set range, we combine the GPR predicted guidance command with an analytical guidance law based on the confidence level.

The analytical guidance law with impact time constraints is given by
\begin{equation}
a_p=-\frac{NV^2}{R}\sin \sigma +\frac{K\phi \left( \sigma \right) \left( 2N-1 \right) V^2}{Rt_{\text{go}}\sin \sigma}\varepsilon _t
\end{equation}
where $N=3$, $K=9$, $\phi \left( \sigma \right) =1-\sqrt{\left| 2\sigma /\pi \right|}$, $t_{\text{go}}=\frac{r}{v}\frac{1+\sin ^2\sigma}{2\left( 2N-1 \right)}$ and $\varepsilon_t=t_f-(t+t_{\text{go}})$.

Considering the posterior distribution of $u^*$ in equation (\ref{posterior distribution}), the confidence of the predicted guidance command is designated as
\begin{equation}
\rho = 1 - \frac{1}{\sqrt{2\pi \sigma_*}}\exp \left( -\frac{\left( a_p-\mu_* \right) ^2}{2\sigma_* ^2} \right) 
\end{equation}

The total guidance command weighted by the confidence level is then expressed as
\begin{equation}
a_n=\rho \mu_* + (1-\rho) a_p
\end{equation}

\begin{figure}[hbt!]
    \centering
    \includegraphics[width=0.9\textwidth]{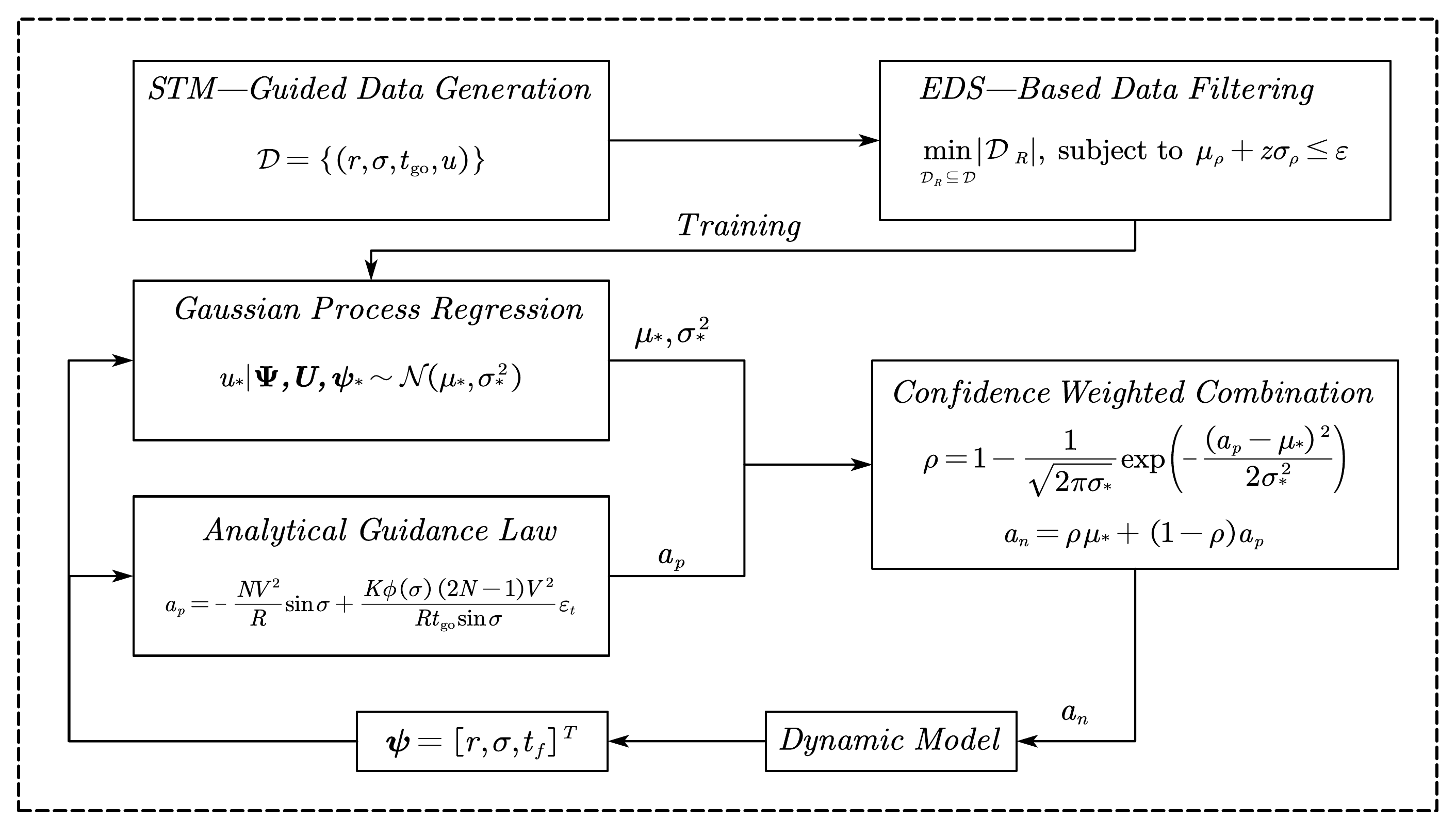}
    \caption{\label{fig: diagram} Diagram of confidence-aware learning framework}
\end{figure}

The diagram of the proposed confidence-aware learning framework is illustrated as Fig. \ref{fig: diagram}. In conclusion, leveraging the proposed data generation and filtering method, a lightweight and representitive dataset is obtained. Then the proposed GPR-based guidance strategy achieves real-time adjustment of the weight of the predicted guidance command according to its confidence level. It identifies whether the learned strategy is effective and reduce the weight of ambiguous guidance commands. In the next section, we will verify the effectiveness of the proposed data generation and filtering method. The reliability of the proposed GPR-based confidence-aware learning method is validated by numerical simulations.

\section{Simulation}
\label{sec:Simulation}
To validate the effectiveness of the proposed data generation, filtering, and confidence-aware learning methods, we conduct a series of simulations in the following subsections. In \ref{sec:dataset related simulations}, we specify a desired data range and implement the data generation and filtering method. The controllability and filtering performance of the proposed methods are validated. In \ref{sec:guidance simulations}, we conduct simulations of the confidence-aware learning method for impact angle control guidance. The effectiveness of the learned strategy is verified and the extended applicability of the confidence weighted combination is demonstrated. 
\subsection{Data generation and data filtering}
\label{sec:dataset related simulations}
In this subsection, we generate optimal trajectories leveraging the proposed STM-guided data generation method. Then we conduct data filtering experiments to obtain the reduced but representative dataset.

\begin{table}[hbt!]
	\caption{\label{tab:param table} Parameters for algorithm 1}
	\centering
	\begin{tabular}{lllccccc}
	  \hline
	  Parameters                 &  & Value                 \\\hline
	  Desired distance range           &  & $r_{\min}=0.8, r_{\max}=1.2$             \\
	  Desired heading error angle range      &  & $\sigma_{\min}=0, \sigma_{\max}=\pi/2$       \\
	  Desired terminal time range &  & $t_{\min}=1.2T, t_{\max}=2T$     \\
	  Distance step                  &  & $\Delta r = 0.05$ \\
	  Heading error angle step  &  & $\Delta \sigma=0.05$              \\
	  Terminal time step          &  & $\Delta t_f=0.04T$              \\
	  Acceleration constraint          &  & $u_m=5$              \\
	  \hline
	\end{tabular}
  \end{table}

The inputs of the algorithm 1 are shown in Table \ref{tab:param table}, where $T$ represents the theoretical minimum of terminal time under acceleration constraints, which can be obtained by 
\begin{equation}
\begin{aligned}
	\alpha &=\arctan \left( \frac{r\cos \sigma -\sqrt{r^2-2rR_{\min}\sin \sigma}}{2R_{\min}-r\sin \sigma} \right)\\
	T&=2R_{\min}\alpha +\frac{r\sin \sigma}{\sin \left( 2\alpha \right)}-R_{\min}\tan \alpha\\
\end{aligned}
\end{equation}
where $R_{\min}$ denotes minimum turning radius $R_{\min} = v^2/u_m = 0.2$.

\begin{figure}[hbt!]
    \centering
	\begin{subfigure}{0.48\textwidth}
		\centering
		\includegraphics[scale=0.4]{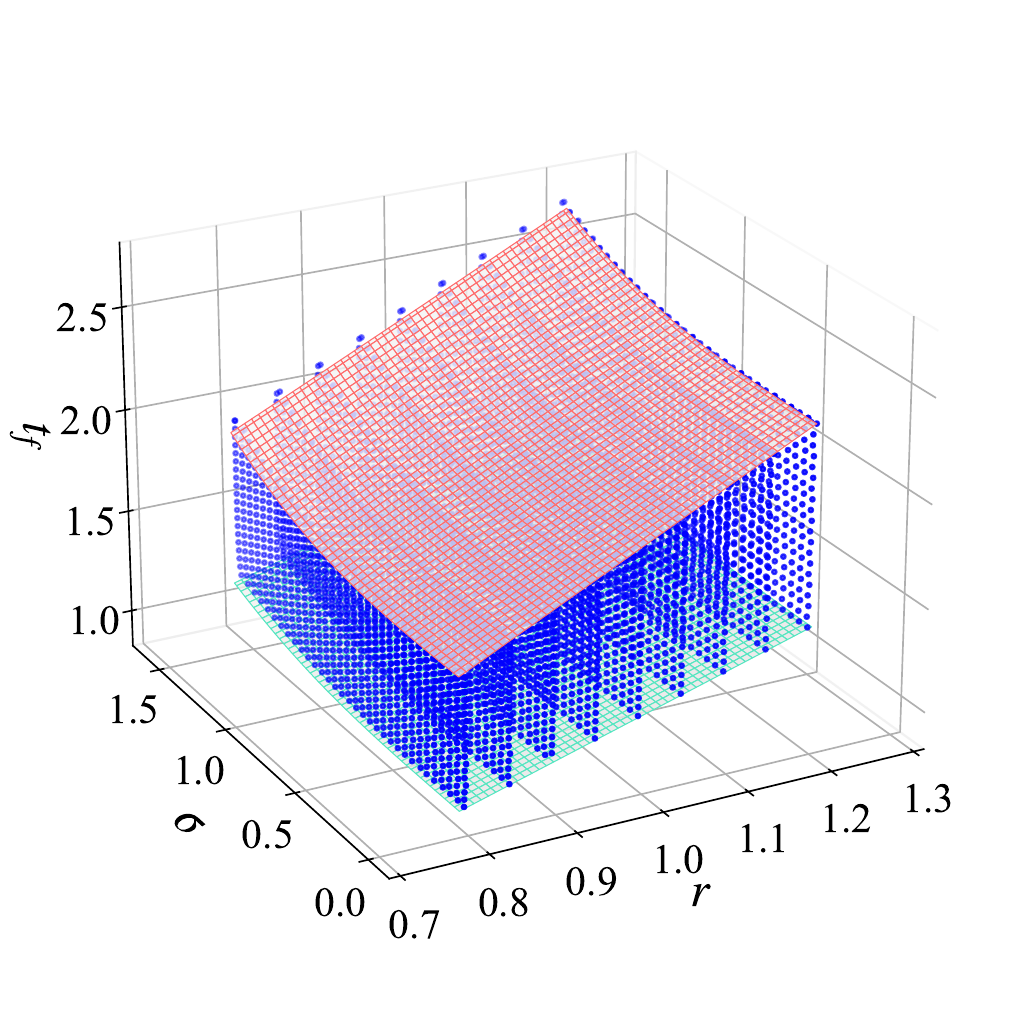}
		\caption{The proposed method}
		\label{fig:InitialState}
	\end{subfigure}
	\hfill
	\begin{subfigure}{0.48\textwidth}
		\centering
		\includegraphics[scale=0.4]{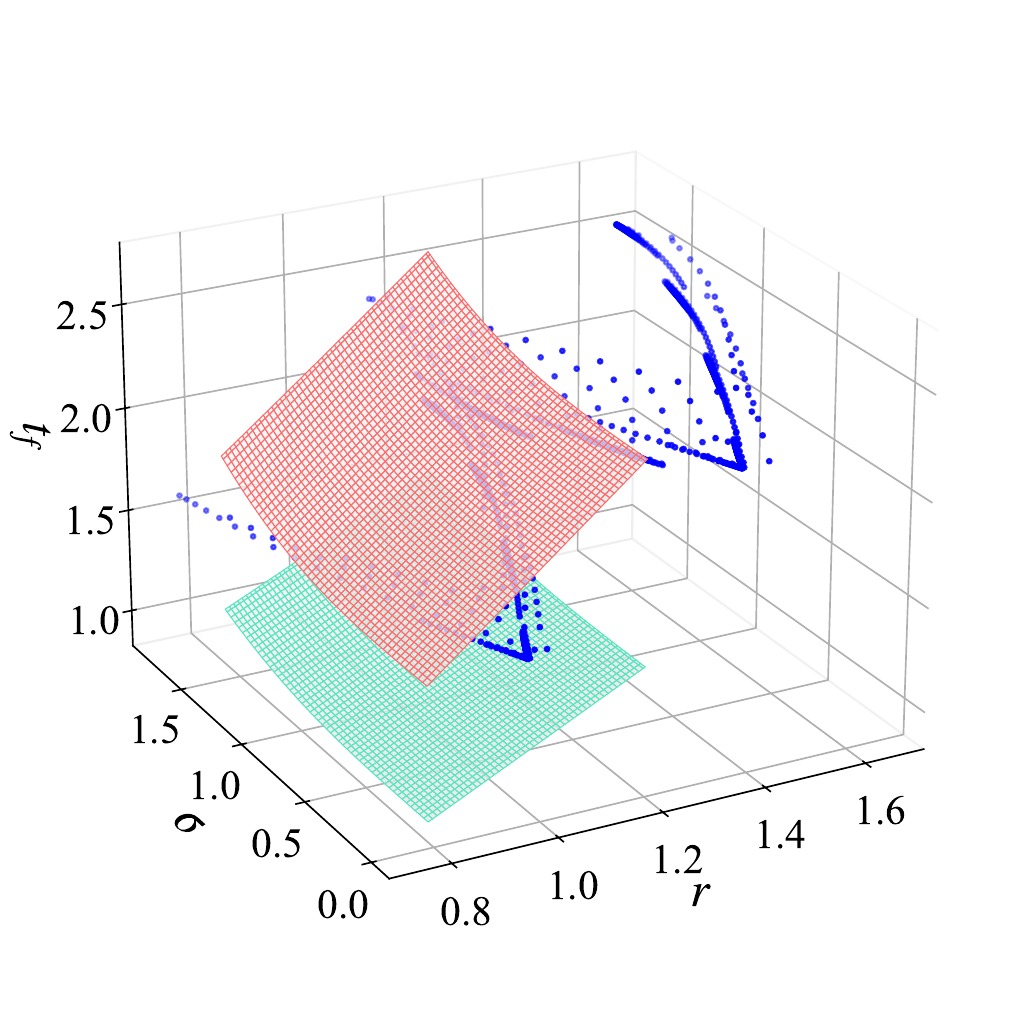}
		\caption{The BGOE method}
		\label{fig:InitialStateUniform}
	\end{subfigure}
	\caption{Comparison of initial state regions}
	\label{fig:RegionComparison}
\end{figure}

Fig. \ref{fig:RegionComparison} exhibits the distribution of initial states of generated optimal trajectories. The green and red surfaces denote terminal time bounds $t_{\min}$ and $t_{\max}$, respectively. Blue dots represent initial states of optimal trajectories. Fig. \ref{fig:InitialState} demonstrates that the proposed data generation method achieves uniform and complete coverage of the desired initial state space. In contrast, when using the BGOE method with uniformly selected terminal costates, initial states of generated optimal trajectories are not controllable, as shown in Fig. \ref{fig:InitialStateUniform}. The proposed region-controllable optimal data generation method helps generate desired optimal trajectories within a specified region, which facilitates subsequent guidance strategy learning.

\begin{figure}[hbt!]
    \centering
	\begin{subfigure}{0.8\textwidth}
		\centering
		\includegraphics[scale=0.43]{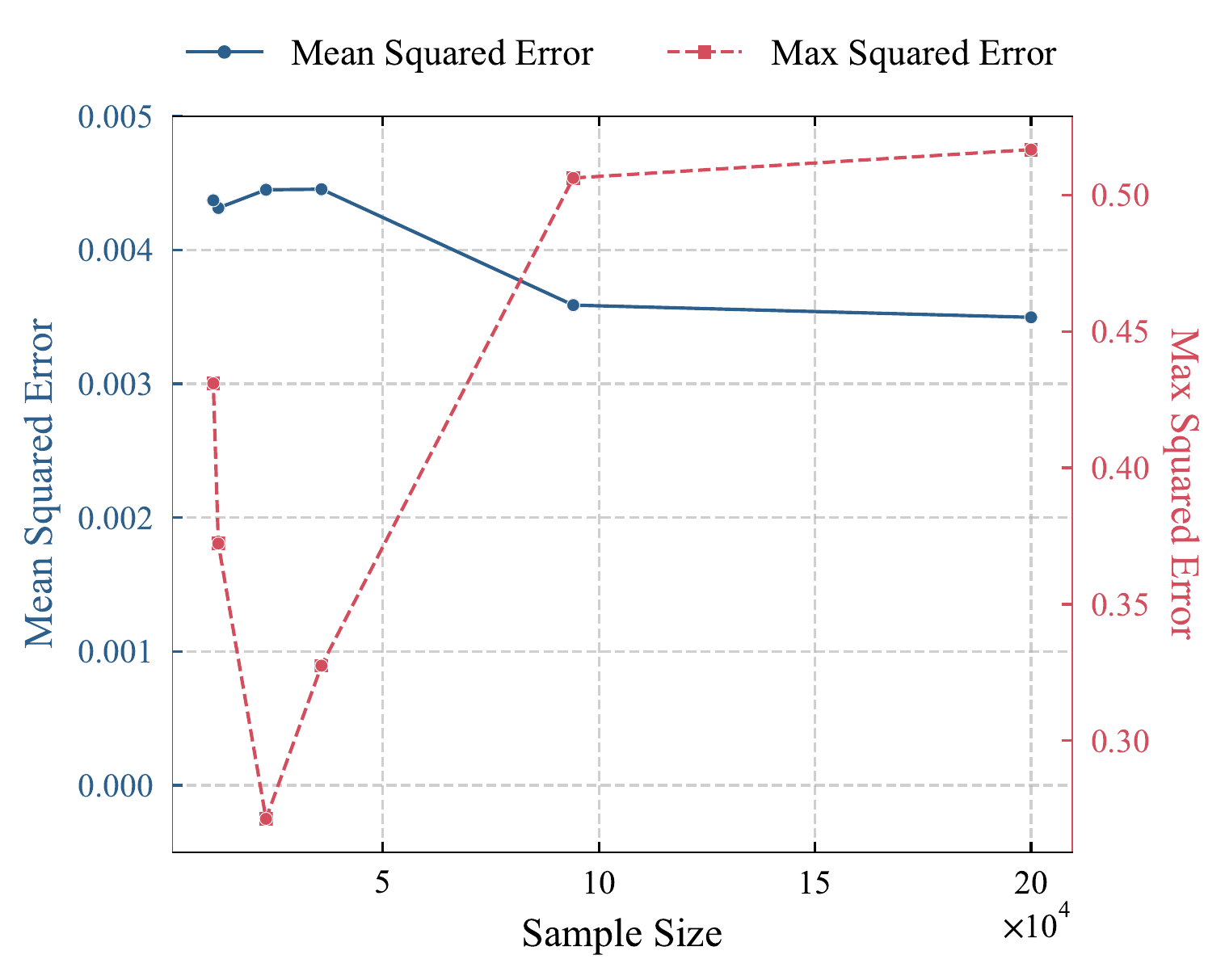}
		\caption{Mean and max squared error with different sample size}
		\label{fig:data_filtering_curve}
	\end{subfigure}
	\vfill
	\begin{subfigure}{0.48\textwidth}
		\centering
		\includegraphics[scale=0.4]{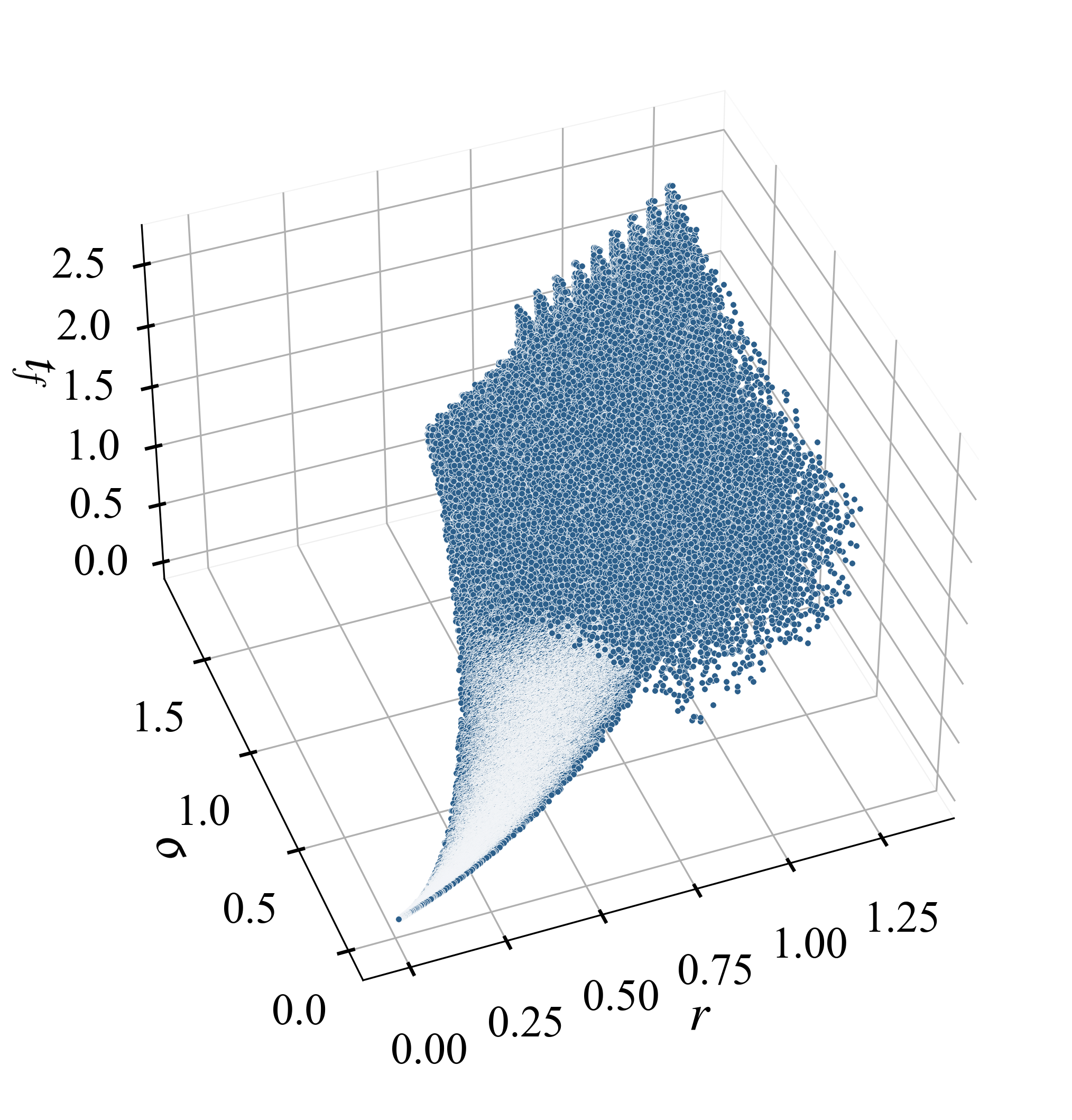}
		\caption{Original samples}
		\label{fig:original_points}
	\end{subfigure}
	\hfill
	\begin{subfigure}{0.48\textwidth}
		\centering
		\includegraphics[scale=0.4]{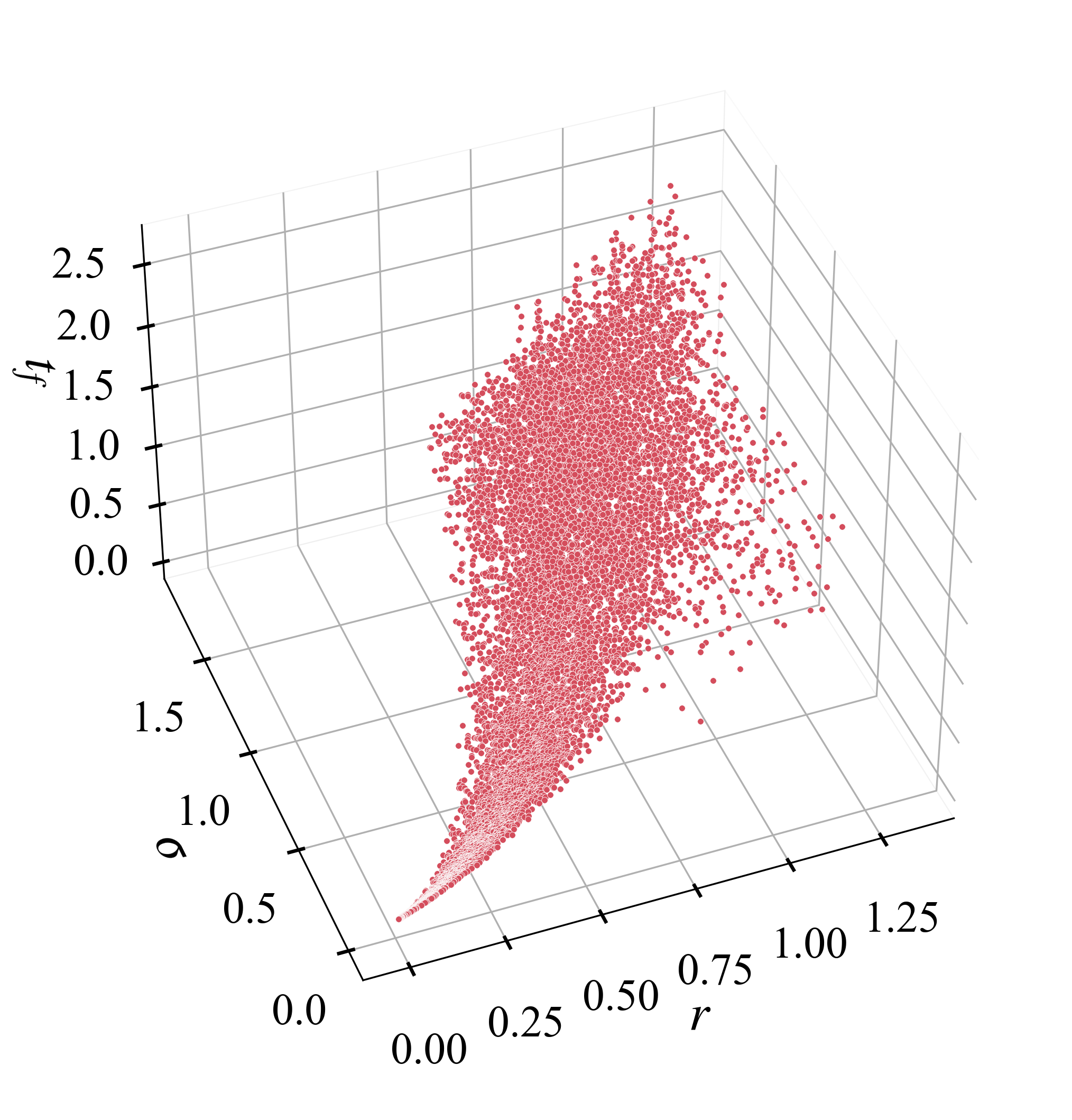}
		\caption{Managed Samples}
		\label{fig:managed_points}
	\end{subfigure}
	\caption{Data filtering by the error distribution smoothing method}
	\label{fig:data_filtering}
\end{figure}

To further reduce the sample size and improve learning efficiency, the error distribution smoothing method is utilized to manage the generated optimal samples. We obtain a series of subsets with different sample sizes by gradually incrementing the threshold $\varepsilon$. Then we train the GPR model on these subsets and test the trained model on the original dataset. Mean squared errors and max squared errors for different sample sizes are illustrated in Fig. \ref{fig:data_filtering_curve}. As the number of samples decreases, the mean squared error of the GPR model slightly increases. However, the max squared error initially decreases and then increases with the reduction in sample size. This phenomenon occurs because the error distribution smoothing method retains more samples in regions with larger errors while minimizing the number of samples in regions with smaller errors. Experimental results demonstrate that applying a threshold of $\varepsilon=0.01$ reduces the subset size to $22,971$ samples, achieving a $90\%$ reduction in data volume while preserving accuracy. Notably, the maximum squared error is significantly reduced, while the mean squared error does not increase substantially. The original sample set and the managed sample set are illustrated in Fig. \ref{fig:original_points} and \ref{fig:managed_points}, respectively. It can be observed that the managed sample set effectively covers the original sample space while significantly reducing the data volume. In the following sections, we will utilize the managed sample set for guidance strategy learning.

\subsection{Impact time control optimal guidance}
\label{sec:guidance simulations}
In this subsection, we verify the effectiveness and reliability of the proposed GPR-based confidence-aware learning method for the impact time control optimal guidance. The GPR model is trained using \textit{GPyTorch} library in Python \cite{gardner_gpytorch_2021}.

\begin{figure}[hbt!]
    \centering
    \begin{subfigure}{0.48\textwidth}
      \centering
      \includegraphics[scale=0.4]{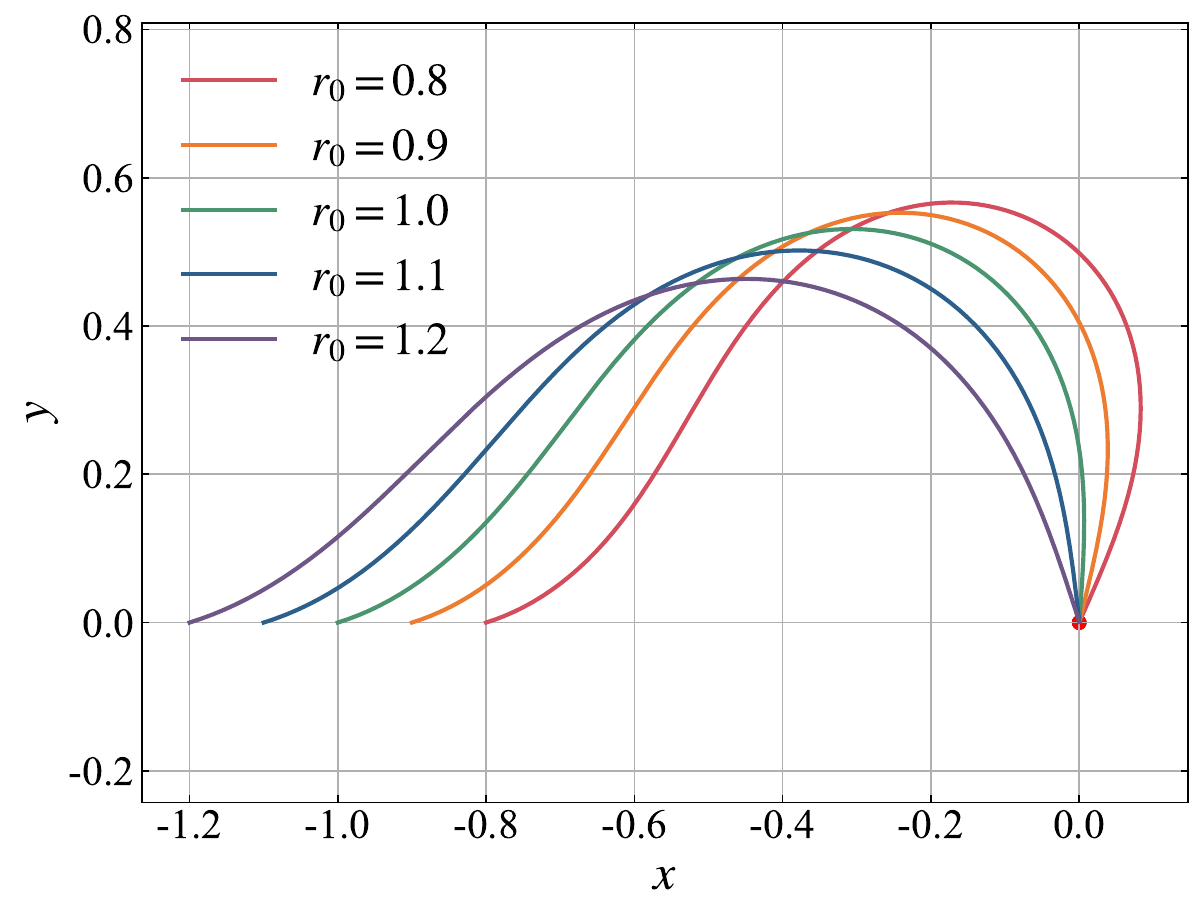}
      \caption{Trajectories}
      \label{fig:Trajectories1}
    \end{subfigure}
    \hfill
    \begin{subfigure}{0.48\textwidth}
	  \centering
      \begin{subfigure}{1\textwidth}
        \centering
        \includegraphics[scale=0.4]{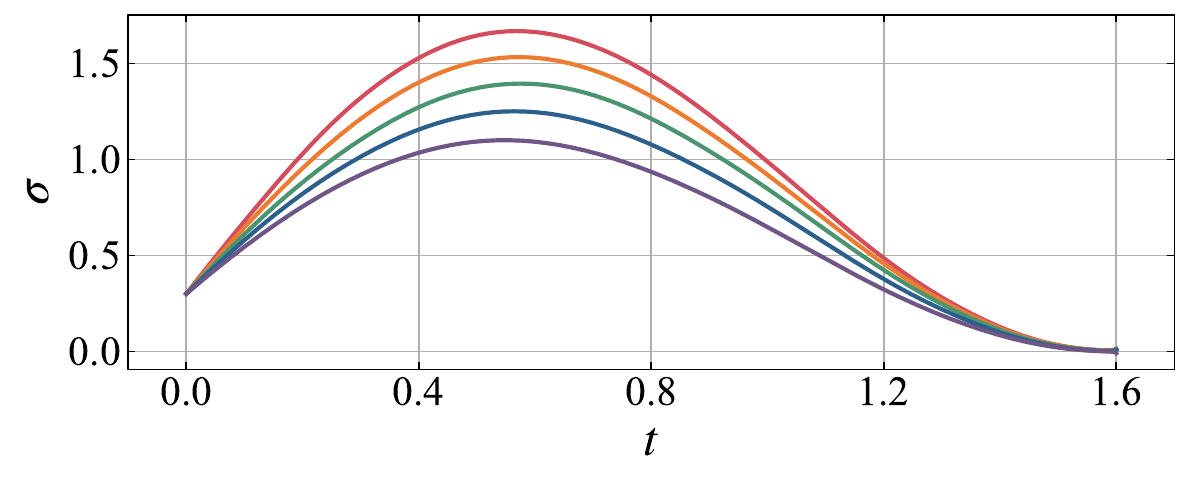}
        \caption{Heading error angle $\sigma$}
        \label{fig:t-sigma1}
      \end{subfigure}
      \begin{subfigure}{1\textwidth}
        \centering
        \includegraphics[scale=0.4]{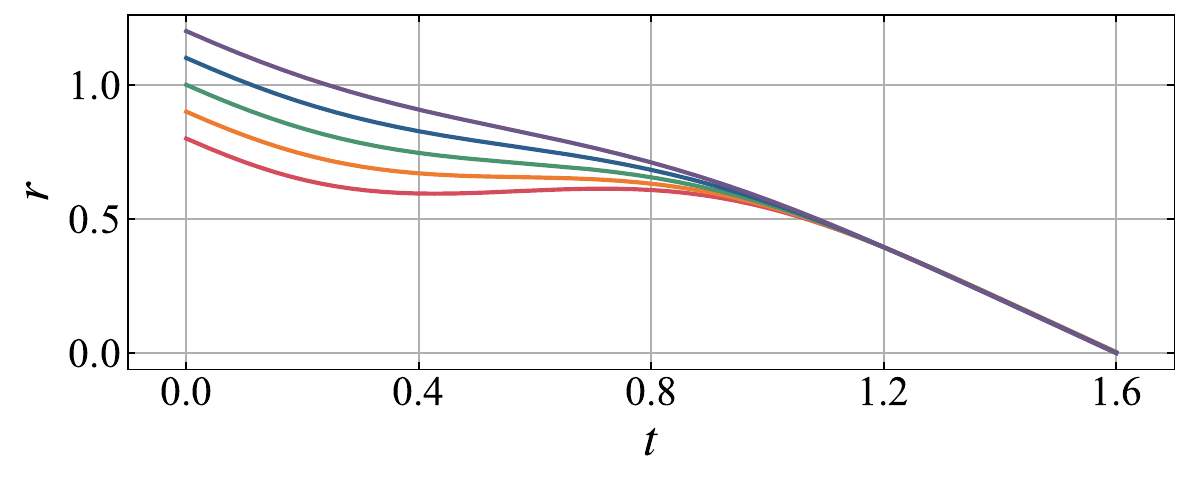}
        \caption{Relative distance $r$}
        \label{fig:t-r1}
      \end{subfigure}
    \end{subfigure}
    \caption{Profiles of impact time control guidance with different $r_0$}
    \label{fig:profiles1}
\end{figure}

\begin{figure}[hbt!]
    \centering
    \begin{subfigure}{0.48\textwidth}
      \centering
      \includegraphics[scale=0.4]{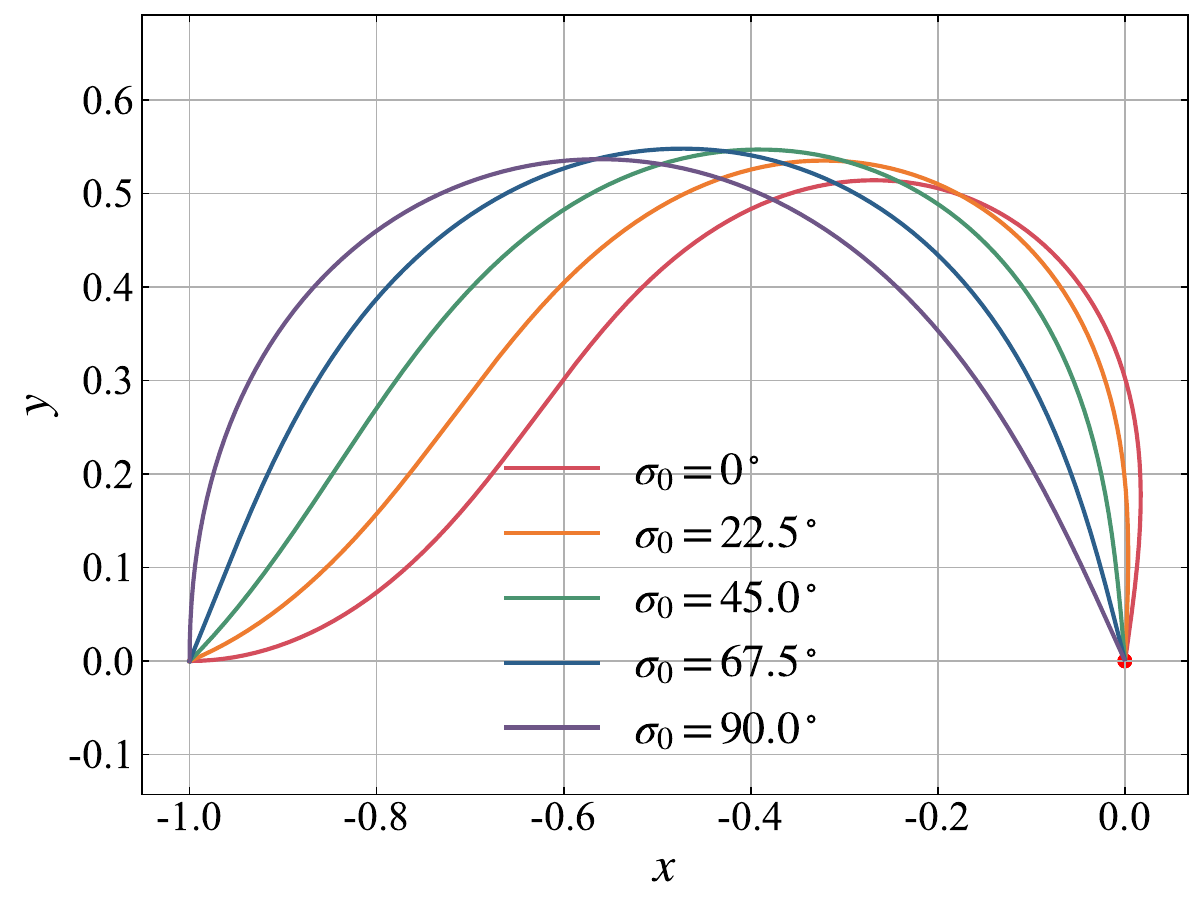}
      \caption{Trajectories}
      \label{fig:Trajectories2}
    \end{subfigure}
    \hfill
    \begin{subfigure}{0.48\textwidth}
	  \centering
      \begin{subfigure}{1\textwidth}
        \centering
        \includegraphics[scale=0.4]{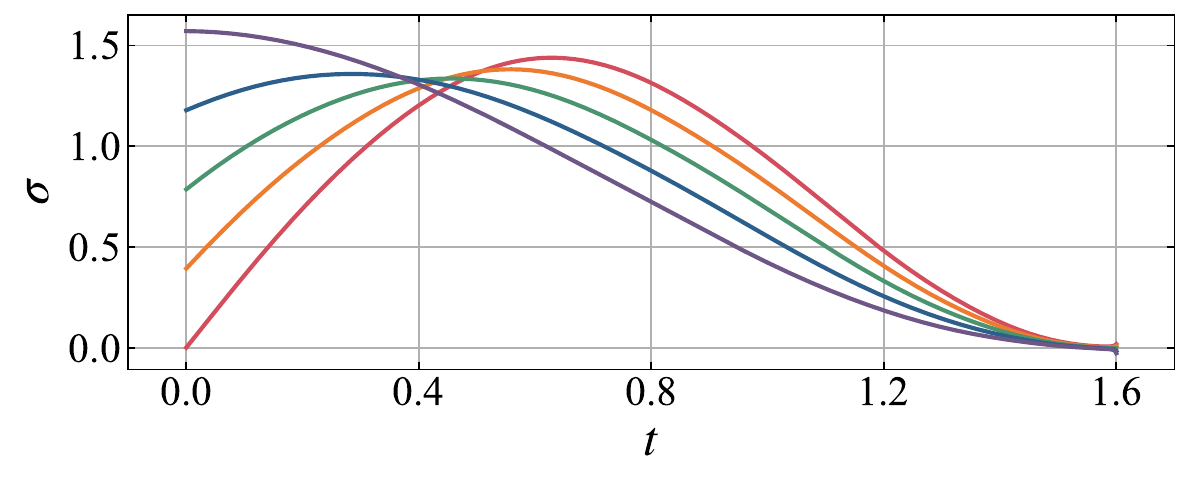}
        \caption{Heading error angle $\sigma$}
        \label{fig:t-sigma2}
      \end{subfigure}
      \begin{subfigure}{1\textwidth}
        \centering
        \includegraphics[scale=0.4]{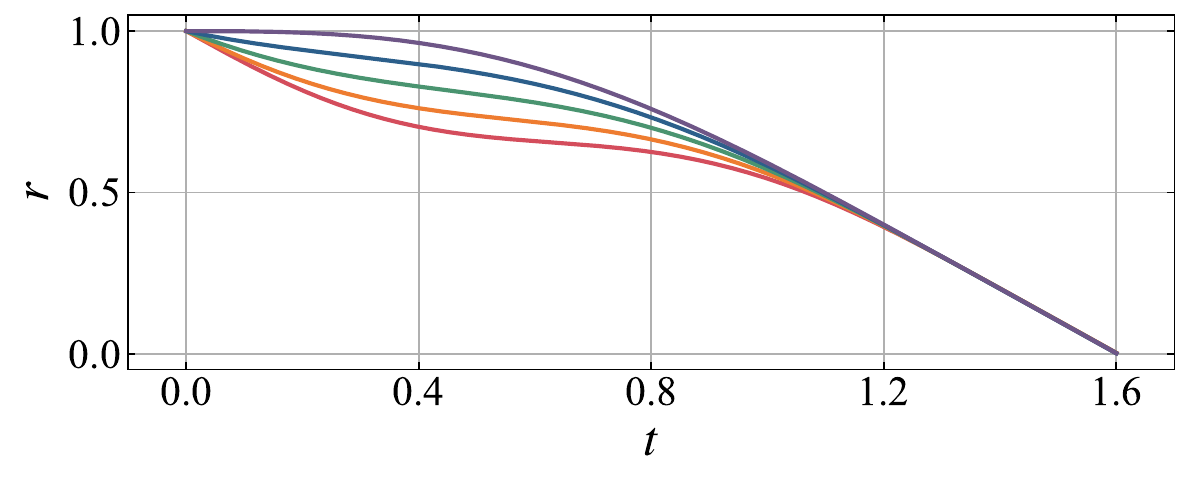}
        \caption{Relative distance $r$}
        \label{fig:t-r2}
      \end{subfigure}
    \end{subfigure}
    \caption{Profiles of impact time control guidance with different $\sigma_0$}
    \label{fig:profiles2}
\end{figure}

\begin{figure}[hbt!]
    \centering
    \begin{subfigure}{0.48\textwidth}
      \centering
      \includegraphics[scale=0.4]{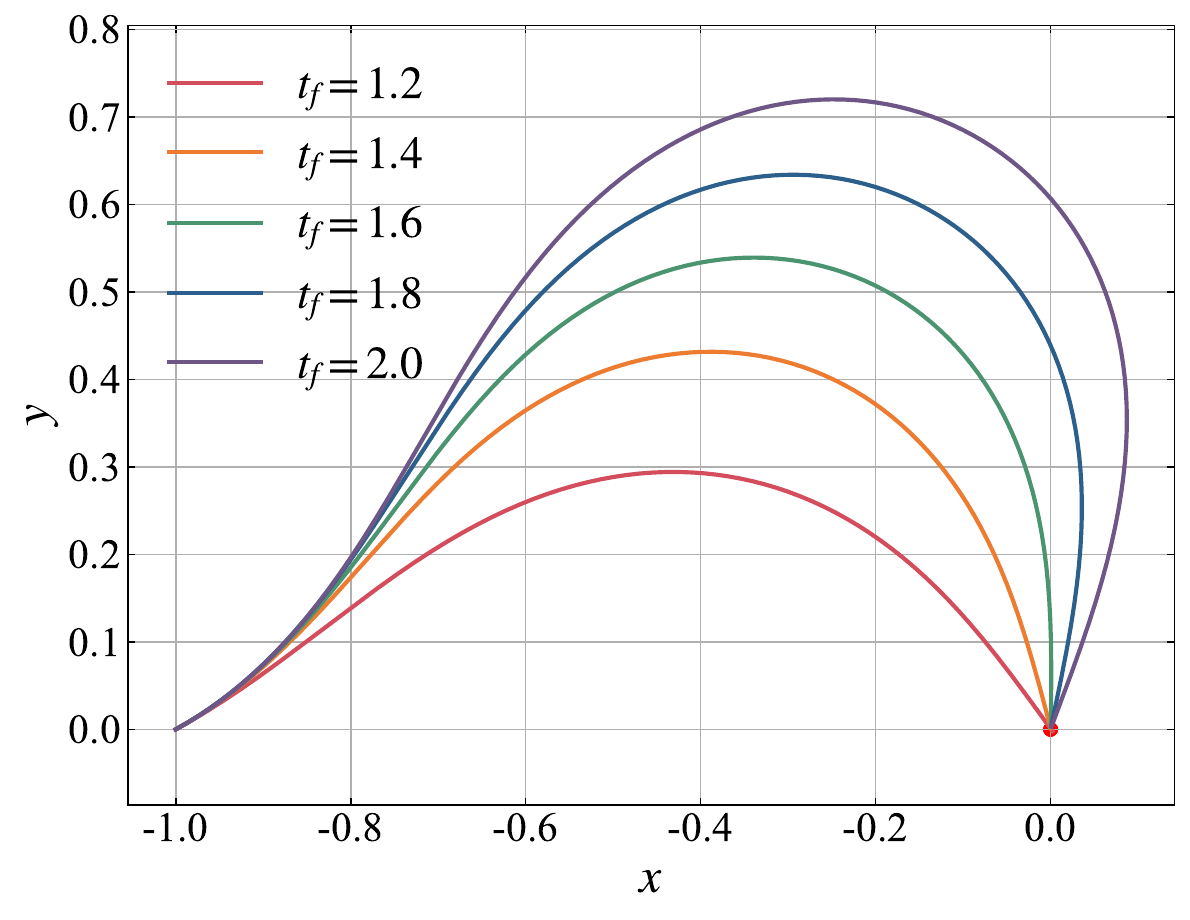}
      \caption{Trajectories}
      \label{fig:Trajectories3}
    \end{subfigure}
    \hfill
    \begin{subfigure}{0.48\textwidth}
	  \centering
      \begin{subfigure}{1\textwidth}
        \centering
        \includegraphics[scale=0.4]{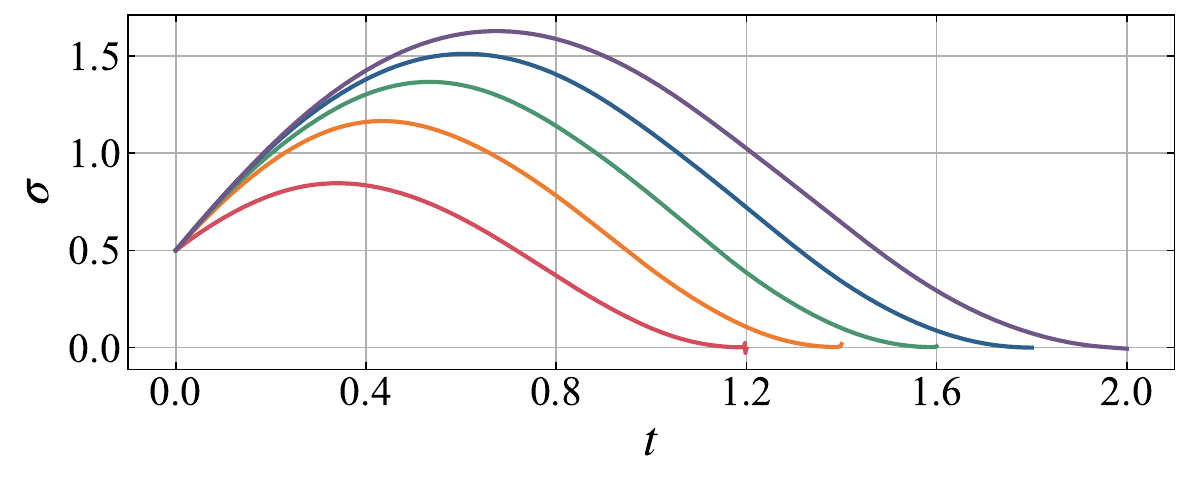}
        \caption{Heading error angle $\sigma$}
        \label{fig:t-sigma3}
      \end{subfigure}
      \begin{subfigure}{1\textwidth}
        \centering
        \includegraphics[scale=0.4]{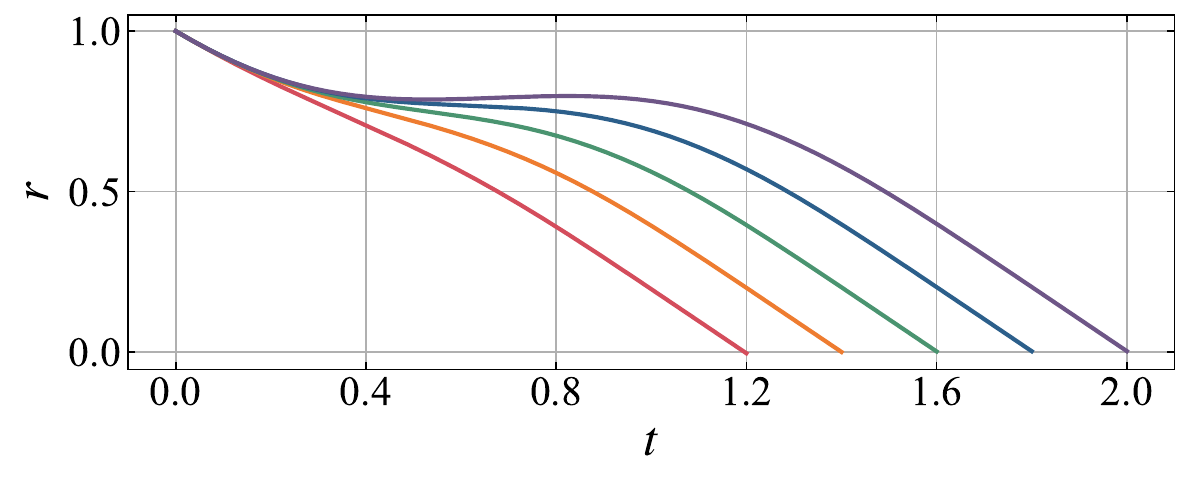}
        \caption{Relative distance $r$}
        \label{fig:t-r3}
      \end{subfigure}
    \end{subfigure}
    \caption{Profiles of impact time control guidance different $t_f$}
    \label{fig:profiles3}
\end{figure}

We conduct simulations for cases of different initial states, including different initial distances $r_0$ varying from from $0.8$ to $1.2$, different initial heading error angles $\sigma_0$ varying from $0$ to $90^\circ$ and different terminal time constraints $t_f$ varying from $1.2$ to $2.0$. Profiles of different guidance cases are presented in Fig. \ref{fig:profiles1}, Fig. \ref{fig:profiles2} and Fig. \ref{fig:profiles3}, respectively. It can be observed that all missiles successfully impact the target at the specified time with terminal heading error angles $\sigma_f=0$ and relative distances $r_f=0$, indicating that the generated and filtered dataset and the GPR-based learning method can achieve satisfying accuracy for constrained guidance problems.

\begin{figure}[hbt!]
\centering
\begin{subfigure}{0.48\textwidth}
	\centering
	\includegraphics[scale=0.4]{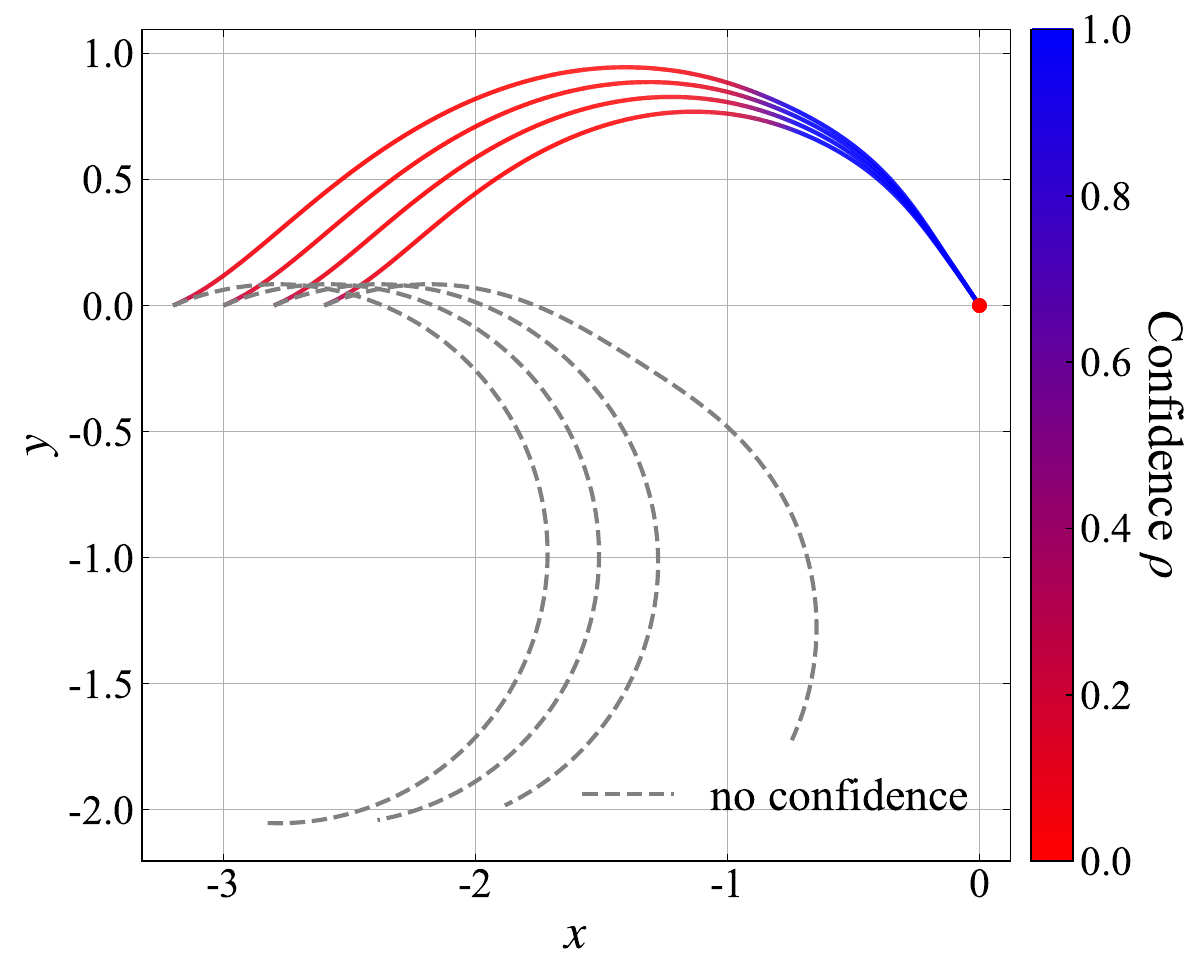}
	\caption{Trajectories}
	\label{fig:Trajectories_Compare}
\end{subfigure}
\hfill
\begin{subfigure}{0.48\textwidth}
	\centering
	\begin{subfigure}{1\textwidth}
	\centering
	\includegraphics[scale=0.4]{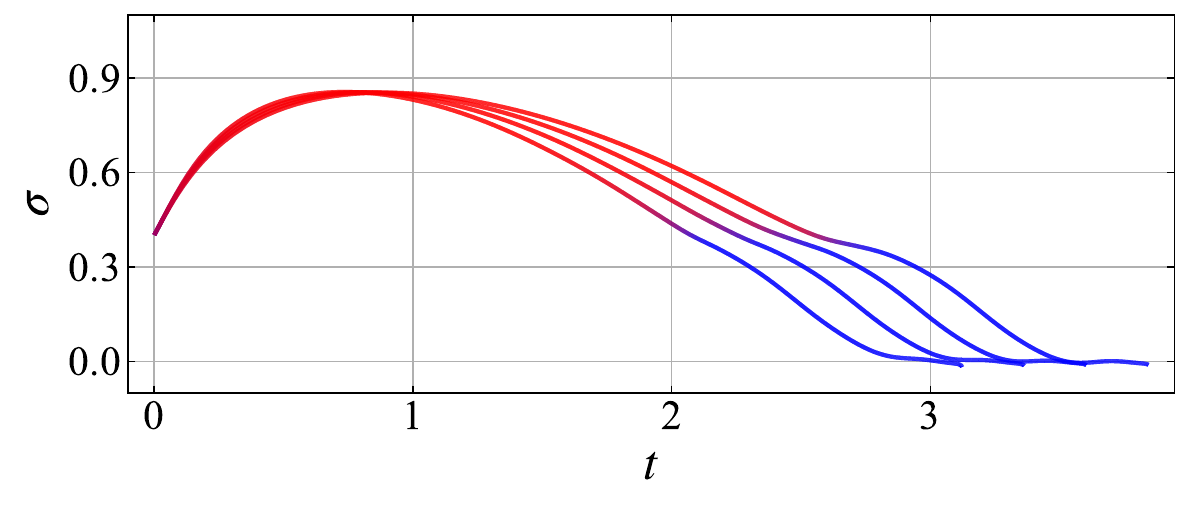}
	\caption{Heading error angle $\sigma$}
	\label{fig:t-sigma_Compare}
	\end{subfigure}
	\begin{subfigure}{1\textwidth}
	\centering
	\includegraphics[scale=0.4]{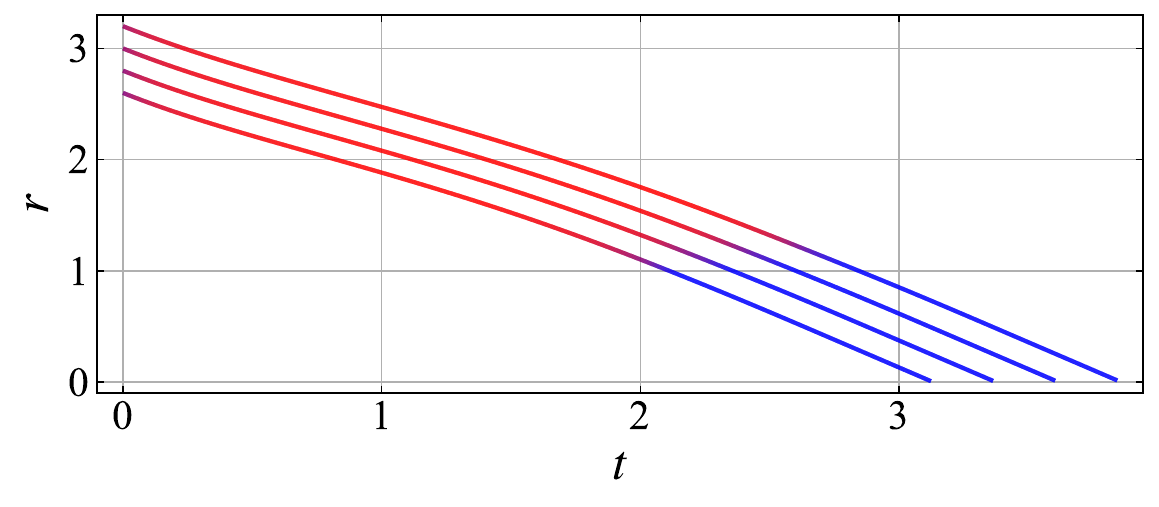}
	\caption{Relative distance $r$}
	\label{fig:t-r_Compare}
	\end{subfigure}
\end{subfigure}
\caption{Profiles of impact time control guidance}
\label{fig:profiles_Compare}
\end{figure}

The reliability and confidence-awareness of the proposed method are further verified in additional simulations with initial states $r_0=\{2.6, 2.8, 3.0, 3.2\}$, $\sigma_0=2.4$ and $t_f=1.2T$, which are far beyond the range of the training set. Guidance profiles for these cases are shown in Fig. \ref{fig:profiles_Compare}. The gray dashed lines in Fig. \ref{fig:Trajectories_Compare} represent trajectories using only GPR-predicted commands. Since it exceeds the range of the training set and the confidence level is not considered, the missile fails to hit the target. The red-to-blue gradient in colored curves indicates the confidence level $\rho$ of the predicted optimal guidance command provided by the GPR model under the current state. As demonstrated in the Fig. \ref{fig:profiles_Compare}, during the initial phase of the guidance process, the confidence level of the GPR's predicted guidance command is relatively low, and the analytical guidance law primarily takes effect. When the missile state enters the range of the training set, the confidence level of the GPR's predicted guidance command increases significantly, and the GPR model plays a dominant role. Ultimately, all missiles successfully hit the target at the designated terminal time. It can be observed that the confidence boundary transition occurs near $r=1.2$, which aligns with the range of the generated optimal trajectories. The effectiveness of the proposed confidence-aware learning is thereby validated.

\section{Conclusion}
\label{sec:Conclusion}
For constrained optimal guidance problems, we propose a confidence-aware learning method including data generation, data filtering and strategy learning. To rapidly generate optimal trajectories within a specified region, we propose a region-controllable data generation method by improving the backward generation of optimal examples method through state-transition-matrix-enhanced backward generation of optimal examples. To reduce sample size while preserving learning accuracy, we conduct data filtering utilizing the error distribution smoothing method. The proposed data generation and filtering method helps construct a high-quality dataset. For strategy learning, a GPR model is trained and combined with an analytical guidance law, enabling real-time confidence evaluation of predicted guidance commands and autonomously adjusting the weight between the learned strategy and the analytical law. Simulation results demonstrate that the proposed confidence-aware learning method can accurately judge when the learned strategy is effective and expand the applicability of the learned strategy. For future research, more refined data generation and learning methods can be explored to broaden the adaptability scope of learning strategies.

\bibliography{references}

\end{document}